\documentclass[showpacs,preprintnumbers,amsmath,amssymb,12pt,pre,aps]{revtex4}
\usepackage{amsmath,amssymb}
\usepackage{epsfig}
\usepackage{citesort}
\usepackage{graphicx}
\usepackage{times}
\newcommand{\greekvektor}[1]{\mbox{\boldmath$#1$\unboldmath}}

\begin{document}

\title{Linear stability,  transient energy growth and the role of viscosity stratification
 in compressible plane Couette flow\footnote{Physical Review E, vol. {\bf 77}, 036322 (2008, March)}
}
\author{M. Malik and  J. Dey}
\affiliation{
Department of Aerospace Engineering, 
Indian Institute of Science, Bangalore 560012, India
}
\author{Meheboob Alam\footnote{Corresponding author.
Email: meheboob@jncasr.ac.in}}
\affiliation{
Engineering Mechanics Unit, Jawaharlal Nehru Center for Advanced
Scientific Research, Jakkur PO, Bangalore 560064, India 
}
                                                                                              
\date{\today}

\begin{abstract}
Linear stability and the non-modal transient energy growth in compressible plane Couette flow
are investigated for two prototype mean flows: (a) the {\it uniform shear} flow
with constant viscosity, and (b) the {\it non-uniform shear} flow with {\it stratified} viscosity.
Both mean flows are linearly unstable for a range of supersonic Mach numbers ($M$). 
For a given $M$, the critical Reynolds number ($Re$) 
is significantly smaller for the uniform shear flow than its non-uniform shear counterpart;
for a given $Re$, the {\it dominant} instability (over all 
stream-wise wavenumbers, $\alpha$) of each mean flow
belongs  different modes for a range of supersonic $M$.
An analysis of perturbation energy reveals that
the instability is primarily caused by an excess transfer of energy from mean-flow to perturbations.
It is  shown that the energy-transfer from mean-flow 
occurs close to the moving top-wall for ``mode I'' instability, whereas it occurs 
in the bulk of the flow domain for ``mode II''. 
For the non-modal transient growth analysis, it is shown that
the maximum temporal amplification of perturbation energy,
$G_{\max}$, and the corresponding time-scale are significantly larger for the
uniform shear case compared to those for its non-uniform counterpart.
For $\alpha=0$, the linear stability operator can be partitioned
into ${\cal L}\sim \bar{\cal L} + Re^2{\cal L}_p$,
and the $Re$-dependent operator ${\cal L}_p$ is shown to have a negligibly small contribution to
perturbation energy which is responsible for the validity of the 
well-known quadratic-scaling law in uniform shear flow: $G(t/{\it Re}) \sim {\it Re}^2$. 
In contrast, the dominance of ${\cal L}_p$ is responsible for the
invalidity of this scaling-law in non-uniform shear flow.
A reduced inviscid model, based on Ellingsen-Palm-type solution, 
has been shown to capture all salient features of transient
energy  growth of full viscous problem.
For both modal and non-modal instability,
it is shown that the {\it viscosity-stratification} of the underlying mean flow
would lead to a delayed transition in compressible Couette flow.
\end{abstract}


\maketitle

\section{{\label{Intro}}Introduction}

The transition to turbulence has its genesis to the growth of small disturbances
in an otherwise undisturbed laminar flow.
Hence, an understanding of different mechanisms  of instability growth is important
to determine the transition scenarios that eventually lead to turbulence in fluids.
The linear stability theory, via the standard normal-mode approach,
is the starting point of such analyzes
which predict the onset of the growth of a small perturbation
imposed on a laminar flow~\cite{DR85}.
However, there are flow configurations that are  stable according to the 
linear stability theory (i.e., the critical Reynolds number
is infinity, ${\it Re}_{\rm cr}=\infty$)
for which the experiments show a finite ${\it Re}_{\rm cr} (<<\infty)$.
It is reasonable to assume that such sub-critical flows
may be destabilized by the nonlinear effects that are neglected in the linear theory.
Is there any linear mechanism that causes an infinitesimally small
perturbation already present in the flow to grow substantially for a short time?
If this is true, then the nonlinearities 
could take over subsequently to trigger a flow-transition.

Indeed, following the seminal work of Orr~\cite{Orr07},
it has recently been realized that one should investigate
the short-time dynamics of the flow without {\it a priori} assuming the
exponential time-dependence for perturbations.
The key idea is that even though each eigenmode decays
in the asymptotic limit ($t\to\infty$) for a {\it stable} flow, a superposition 
of such stable eigenmodes has potential for large {\it transient} energy
growth before they can be stabilized by the viscosity.
Such transient growth analyzes 
\cite{Gust91,BF92,TTRD93,RSH93,GG94,HSH96,SE97,FI00,KL00,PH06}
have revealed that a flow can sustain large amplification of perturbation energy
even if the flow is linearly stable.
In mathematical terms, the underlying linear stability operator is 
non-normal (non-self-adjoint)~\cite{BF92,TTRD93,Schmid07}
which is responsible for transient energy growth.
This is a possible route to flow transition 
for subcritical flows which has become  an active field of research
during the last ten years~\cite{Schmid07,ESHW07,Gross00}.

It is  known that small changes in the mean-flow can
be stabilizing or destabilizing which is 
an attractive avenue from the viewpoint of controlling or manipulating instabilities.
A recent work~\cite{BCL03} clearly underscores the effects of mean-flow variation
on the stability of {\it incompressible} plane Couette flow-- using the
concept of pseudo-spectra~\cite{TTRD93,RSH93}, these authors
showed that relatively small changes in the mean flow could be {\it destabilizing}.
Another important issue in stability research is the possible 
role of viscosity stratification on instabilities
which has a stabilizing effect,  leading to  a delayed transition.
In incompressible non-Newtonian fluids, the role of viscosity-stratification
in delaying transition is currently debated for which we refer
to a recent work~\cite{NBB07}.

The above issues have not been investigated for {\it compressible} fluids
in a systematic manner.  In this paper, the linear stability characteristics
and the non-modal transient energy growth in the compressible plane Couette flow
are analyzed for two prototype model problems:
(a) uniform shear flow with constant viscosity, and
(b) nonuniform shear flow with stratified viscosity.
The  {\it first} goal of the present work is to understand the similarities and differences 
of the modal and nonmodal stability characteristics between 
these two closely related mean flows of a compressible fluid.
The {\it second} goal is to reveal the role of {\it viscosity-stratification}
on instabilities in a {\it compressible} fluid since we have two prototype
mean-flow configurations in which one has a {\it stratified} viscosity 
across the channel and the other has a constant viscosity.
The {\it third} goal is to characterize the underlying instability
mechanism in compressible Couette flow via an energy analysis.

This paper is organized as follows.
The governing equations and the mean flow are briefly described  in Section~II.
The linear stability problem is formulated in Section~III,
and the related  results are presented in Section~III.
The results on the transient energy growth 
are presented  in Section~IV. 
The summary and conclusions are provided in Section~V.

\section{{\label{EOM}}Equations of Motion and Mean Flow}

Consider a perfect gas of density $\rho^*$ and temperature $T^*$
between two walls that are separated by a distance $h^*$:
the top wall moves with a velocity $U_1^*$ and the lower wall is stationary,
with the top-wall temperature being maintained at $T_1^*$;
here the superscript $*$ denotes dimensional quantities,
and the subscript $1$ refer to the quantities at the top wall.
Let $u^*$, $v^*$ and $w^*$ be the velocity components in the
streamwise ($x^*$), wall-normal ($y^*$) and spanwise ($z^*$) directions, respectively.
The conservation equations for the mass, momentum 
and energy, in dimensionless  form, are:
\begin{eqnarray}
 \frac{\partial \rho}{\partial t} &=& - {\bf \nabla \cdot} (\rho {\bf u})  ,
  \label{cont_eqn} \\
\rho \frac{{\rm D}u_i}{{\rm D}t} &=&
- \frac{1}{\gamma M^2}\frac{\partial p}{\partial x_i}+\frac{1}{{\it Re}} \left[
  \mu\nabla^2u_i + \frac{\partial}{\partial x_i}(\lambda{\bf \nabla\cdot u})
 \right. \nonumber \\
  &+ & 
 \left. 
   \mu\frac{\partial}{\partial x_i}({\bf \nabla\cdot u})
   + (\nabla\mu)\cdot(\nabla u_i) + (\nabla\mu)\cdot\frac{\partial {\bf u}}{\partial x_i}
\right]\\
\rho \frac{{\rm D}T}{{\rm D}t} 
 &=& (1-\gamma)p{\bf \nabla \cdot u} + \frac{\gamma}{\it Re}{\bf\nabla} \cdot 
   \left( \frac{\mu}{\sigma} {\bf\nabla}T \right) + {\Phi}
\label{energy_eqn}
\end{eqnarray} 
with $D/Dt=(\partial/\partial t + {\bf u}\cdot\nabla)$ being the
material derivative, $\Phi$ the dissipative shear work,
and the equation of state is that of a perfect gas: $p = \rho T$.
We have used the separation between the two walls $h^*$ as
the length scale, the top wall velocity, $U_1^*$, and temperature, $T_1^*$,
as the velocity and temperature scale, respectively,
and the inverse of the overall shear rate, $U_1^*/h^*$, as the time scale.
The nondimensional control parameters are the Reynolds number $Re$,
the Prandtl number $\sigma$ and the Mach number $M$:
\begin{equation}
{\it Re} = \frac{\rho^*_1 U^*_1 h^*}{\mu^*_1}, \quad  \ 
 \sigma = \frac{\mu^*c_p^*}{\kappa^*} \quad
  \ \mbox{and} \quad \  M = \frac{U_1^*}{\sqrt{\gamma R T_1^*}}.
\end{equation}
Here $\mu^*$ is the shear viscosity, $\zeta$ the bulk viscosity,
$\kappa^*$ the thermal conductivity, 
$\gamma=c_p^*/c_v^*$ the ratio of specific heats, 
$R$ the universal gas constant and $\lambda= \zeta - 2\mu/3$. 
The bulk viscosity is assumed to be zero
(i.e., $\zeta=0$) such that $\lambda= - 2\mu/3$ (Stokes' assumption).
For all calculations below, $\sigma =0.72$ and $\gamma=1.4$.

\subsection{Constant viscosity: Uniform shear flow}
\label{meanflowseclab}

For the unidirectional steady and fully developed mean flow,
the continuity and the $z$-momentum equations are trivially satisfied.
From the $y$-momentum equation,
it is straightforward to verify that the pressure, $p_0=\rho_0(y)T_0(y)$,
is a constant, which is normalized such that  $p_0=1$. 
(The subscript $0$ is used to designate the mean flow quantities.)
The boundary conditions on the stream-wise velocity $U_0(y)$
and temperature $T_0(y)$ are
\begin{equation}
  U_0(0) = 0,
    \quad U_0(1) = 1, 
    \quad T_0(0) = T_w, 
    \quad  T_0(1) = 1,
\label{eqn_BC1}
\end{equation} 
with $T_w$ being the temperature of the lower wall.

For the constant viscosity model ($\mu_0 = \mbox{constant}$),
the stream-wise velocity varies linearly with $y$:
\begin{equation}
   U_0(y) = y,
\end{equation}
i.e., the shear-rate is {\it uniform}.
It is straightforward to verify that the temperature varies quadratically
with $y$:
\begin{equation}
   T_0(y) = T_r\left[r+(1-r)y-\left(1-\frac{1}{T_r}\right)y^2\right],
\end{equation}
where $T_r$ is the recovery temperature,
\begin{equation}
  T_r = 1+ \frac{(\gamma-1)\sigma M^2}{2},
\end{equation}
and $r = T_w/T_r$ the temperature ratio. 
Note that $r = 1$ corresponds to an adiabatic lower wall.

\subsection{Viscosity stratification: Nonuniform shear flow}

For a temperature-dependent viscosity model, for example, with
Sutherland's law,
\begin{equation}
  \mu(T) = \frac{T^{3/2}(1+C)}{(T+C)}, 
 \quad \mbox{with} \quad C=0.5,
\label{eqn_viscosity}
\end{equation}
the streamwise velocity has a non-uniform shear rate.
For this case, the mean flow equations,
\begin{eqnarray}
  \frac{\rm d}{{\rm d}y}\left(\mu(T_0)\frac{{\rm d}U_0}{{\rm d}y}\right) &=& 0,\nonumber \\ 
  \sigma^{-1}\frac{\rm d}{{\rm d}y}\left(\mu\frac{{\rm d}T_0}{{\rm d}y}\right) 
    + (\gamma-1)M^2\mu\left(\frac{{\rm d}U_0}{{\rm d}y}\right)^2 &=& 0,
\end{eqnarray}
with boundary conditions (\ref{eqn_BC1})
have been solved numerically using the 4th-order Runge-Kutta method.

In contrast to the constant viscosity model, 
for this model the viscosity varies across the channel width, i.e., the mean flow
is characterized  by a {\it stratified} viscosity. 
It is straightforward to verify that  the viscosity at the lower wall 
increases with increasing Mach number, and hence the degree of viscosity stratification
increases with increasing $M$.

\section{{\label{LSA}}Linear Stability Analysis}

For the linear stability analysis,
the mean flow, ${\bf q_0}=(U_0, 0, 0, \rho_0, T_0)^T$,
is perturbed with small amplitude perturbations $\bf q = q_0 + \hat{q}$, 
and the governing equations~(\ref{cont_eqn}) to~(\ref{energy_eqn}) 
are linearized around the mean flow.
Seeking normal mode solutions of the resulting linearized partial 
differential equations,
\begin{equation}
  {\bf \hat{q}}(x,y,z,t) = {\bf q'}(y) \ \exp \ [{\rm i}(\alpha x + \beta z - \omega t)], 
\end{equation}
we obtain a differential eigenvalue system:
\begin{equation}
 {\cal L} {\bf q'} = \omega {\rm I}{\bf q'},
\label{eigeqn}
\end {equation}
where $\cal L$ is the linear stability operator,
${\bf q'} = \{u',v',w',\rho',T'\}^{\bf \rm T}$ is the eigenfunction and 
${\rm I}$ the identity matrix. 
Here $\alpha$ and $\beta$ are the stream-wise and span-wise wave-numbers, respectively, 
and $\omega=\omega_r + {\rm i}\omega_i$ is the complex frequency;
the phase speed of perturbation is given by  $c_r=\omega_r/\alpha$
and the growth/decay rate by $\omega_i$.

The boundary conditions on perturbation variables are taken to be: 
\begin{equation}
\begin{array}{lcl}
    u'(0) &=& 0 = u'(1) \\
   v'(0) &=& 0 = v'(1) \\
   w'(0) &=& 0 = w'(1) \\
    T'(1) &=& 0 = \frac{{\rm d}T'}{{\rm d}y}(0).
\end{array}
\label{eigeqn_bc}
\end{equation}
The Chebyshev spectral method~\cite{MAD06} is used to discretize
the differential eigenvalue problem (\ref{eigeqn}--\ref{eigeqn_bc})
at  $(N+1)$ Gauss-Lobotto collocation points, 
where $N$ is the degree of the Chebyshev polynomial.
This yields an algebraic eigenvalue system, $AX=\omega BX$, which is 
then solved using the QR-algorithm of the Matlab software.

\begin{figure}[h!]
\includegraphics[width=7.0cm]{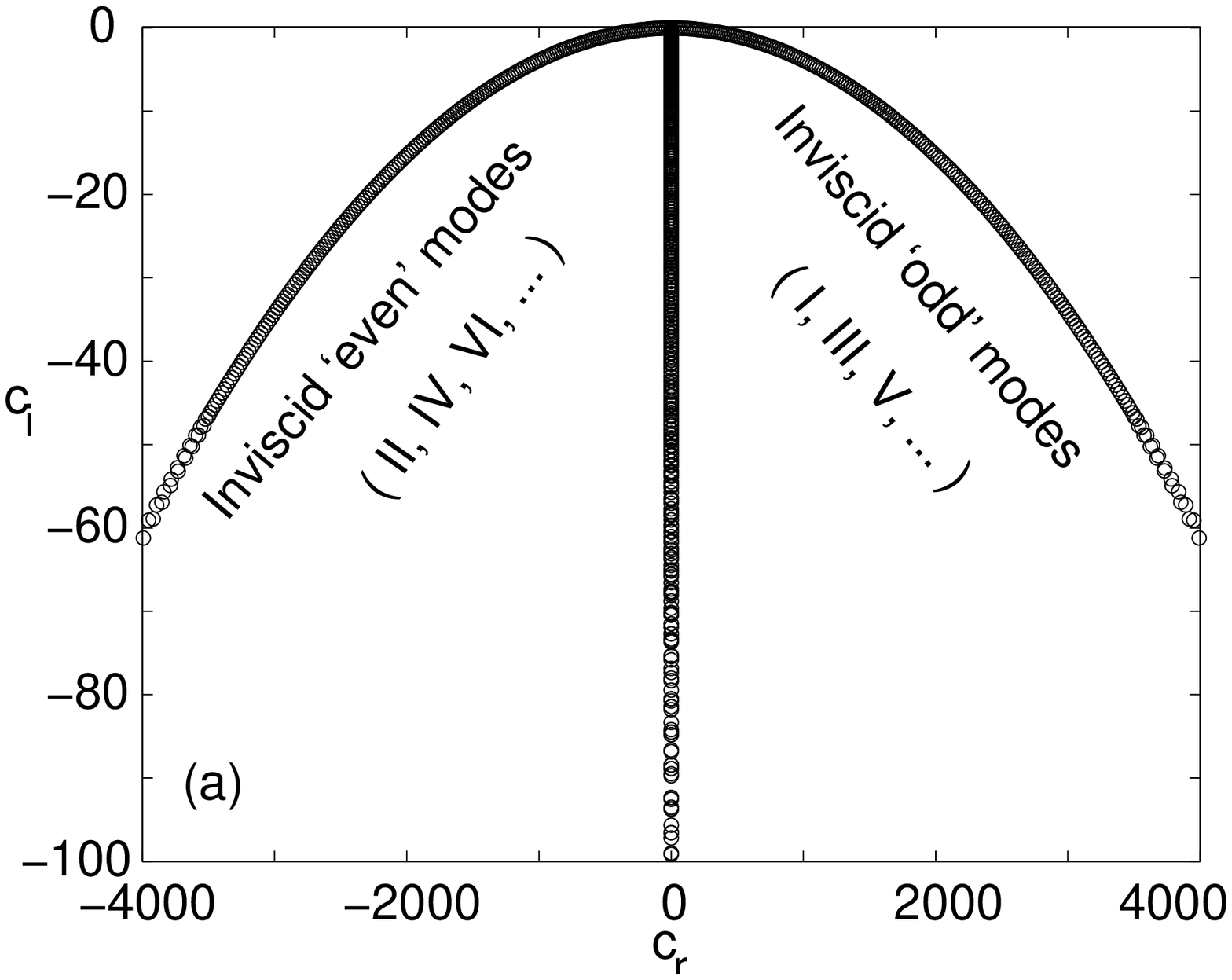}\\
\includegraphics[width=7.0cm]{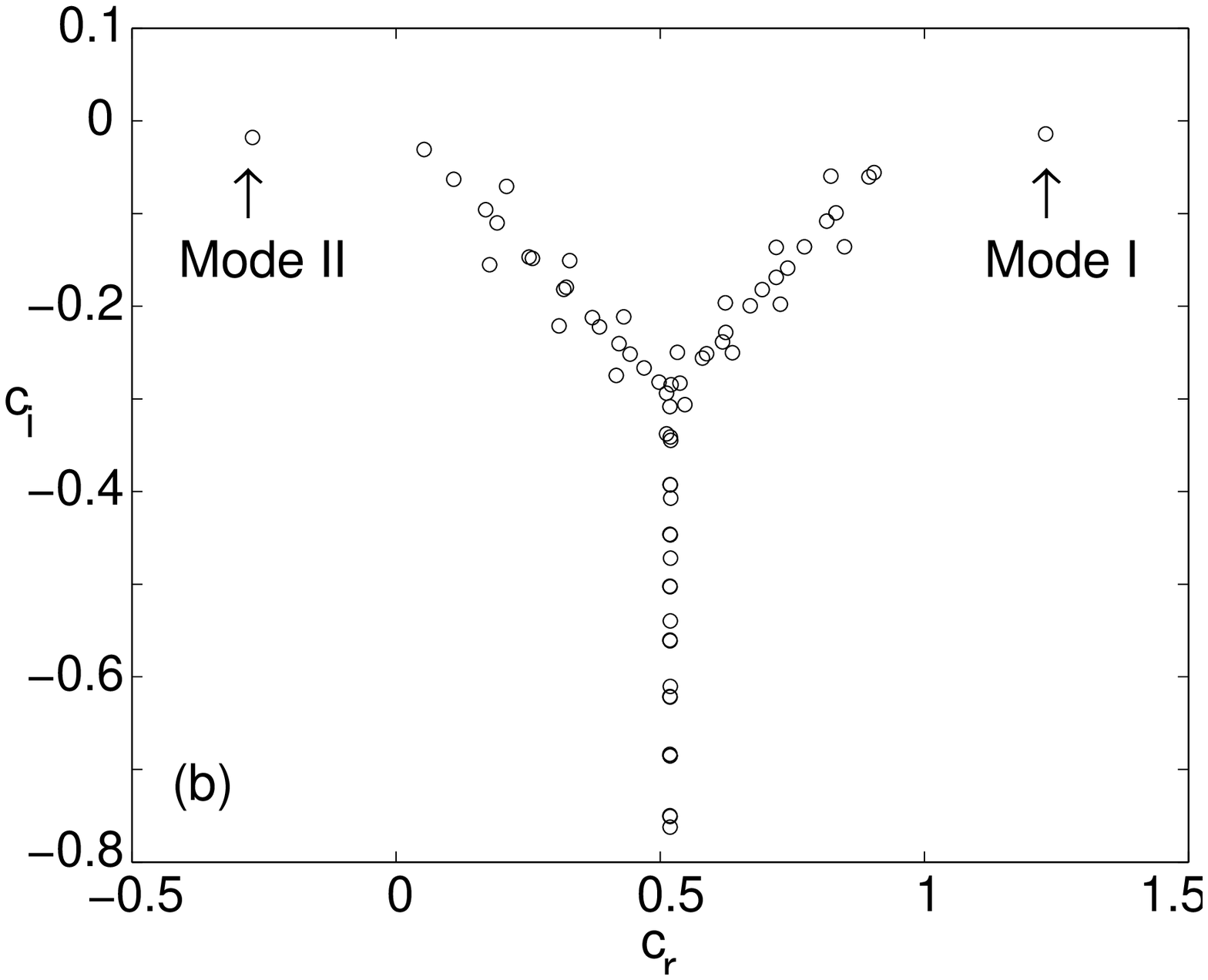}\\
\caption{
Distribution of eigenvalues ($c\equiv \omega/\alpha=c_r + {\rm i}c_i$)
in the complex plane for uniform shear flow
with $Re=10^5$, $M=2$, $\alpha=0.1$ and $\beta=0$.
Panel (b) is the zoom of the viscous modes in panel (a).
According to the phase-speed based classification
of inviscid modes, the mode-III stays  on the right of the mode-I,
the mode-IV is on the left of the mode-II, and so on (see text for details).
}
\label{fig:fig1}
\end{figure}

\subsection{{\label{LSResult}}Spectra and acoustic modes}

Figure~1(a) shows the distribution of eigenvalues,
$c = \omega/\alpha =c_r + {\rm i} c_i$, in the complex plane, and the zoom of Fig.~1(a),
portraying the well-known `Y'-branch of the viscous modes, is shown in Fig.~1(b).
The parameter values are set to $Re=10^5$, $M=5$, $\alpha=0.1$ and $\beta=0$, with $N=150$.
The classification of inviscid eigenvalues (i.e., {\it acoustic} modes)
into {\it odd}- and {\it even}-families in Fig.~1(a)
is based on  their phase speeds~\cite{DEH94}: the {\it odd-modes} (I, III, ...) 
have phase speeds greater than unity in the limit of $\alpha\to 0$,
and the {\it even-modes} (II, IV, ...)
have phase speeds  less than zero  as $\alpha\to 0$.
(Recall that the non-dimensional velocity of the top and bottom
walls are 1 and 0, respectively.)
With increasing $\alpha$, however, the phase speeds of even/odd modes
increases/decreases (not shown), and these modes become unstable
once they enter the viscous range of the spectra (i.e., for $0<c_r<1$)
for a range of supersonic Mach numbers and above some critical value
of Reynolds number (see below).
More specifically, the phase speed of mode I decreases below unity
and that of mode II increases above zero, when they 
degenerate into unstable modes with increasing $\alpha$.
This overall scenario of modal-structure holds for both mean flows;
however, there are important differences 
with regard to the unstable zones in different control
parameter space, the dominant instability
and the critical Reynolds number as detailed below.

\subsection{{\label{DomIns}}Stability diagram and dominant instability}

\begin{figure}[h!]
\includegraphics[width=4.1cm]{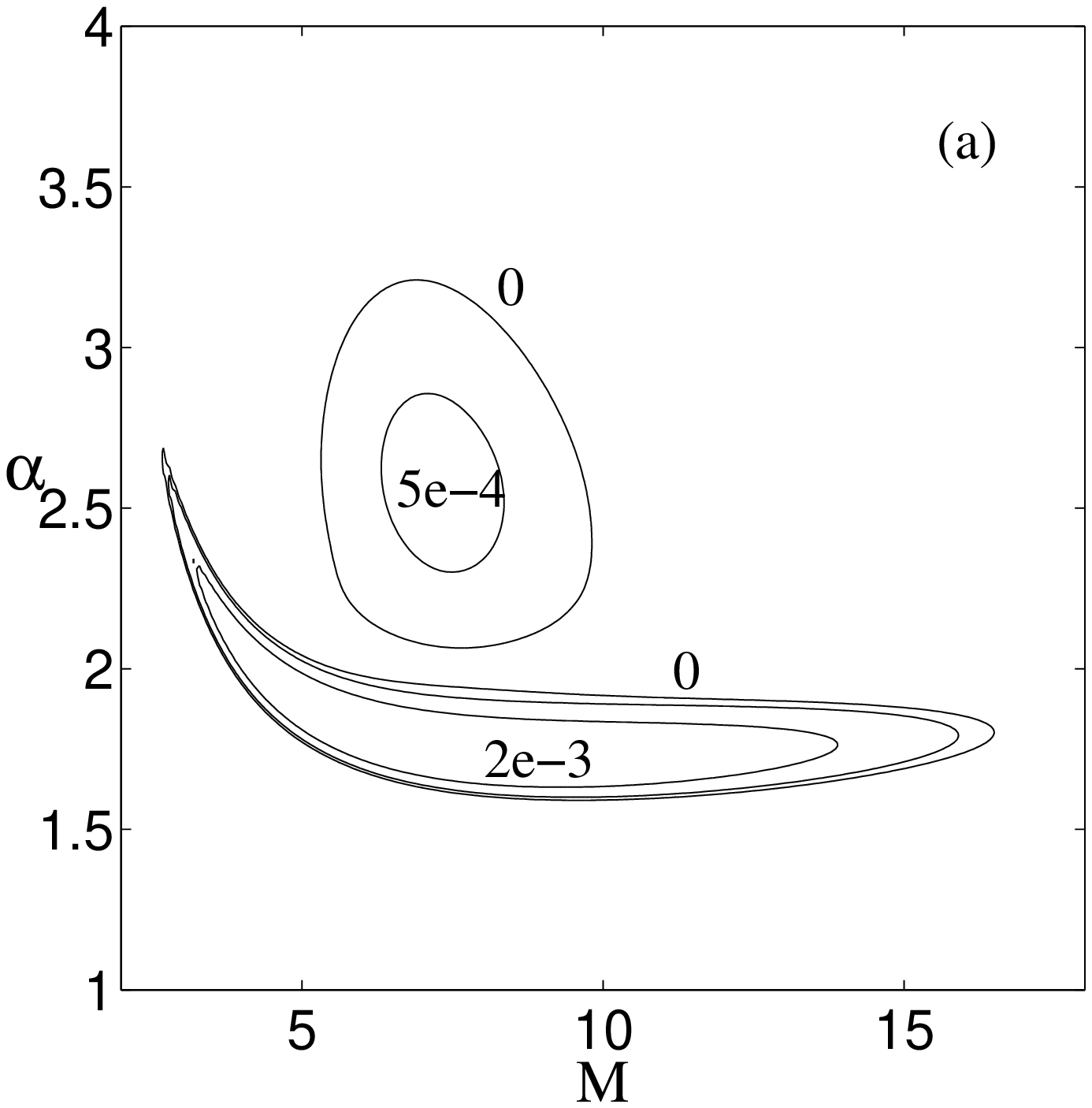}\;\;
\includegraphics[width=4.1cm]{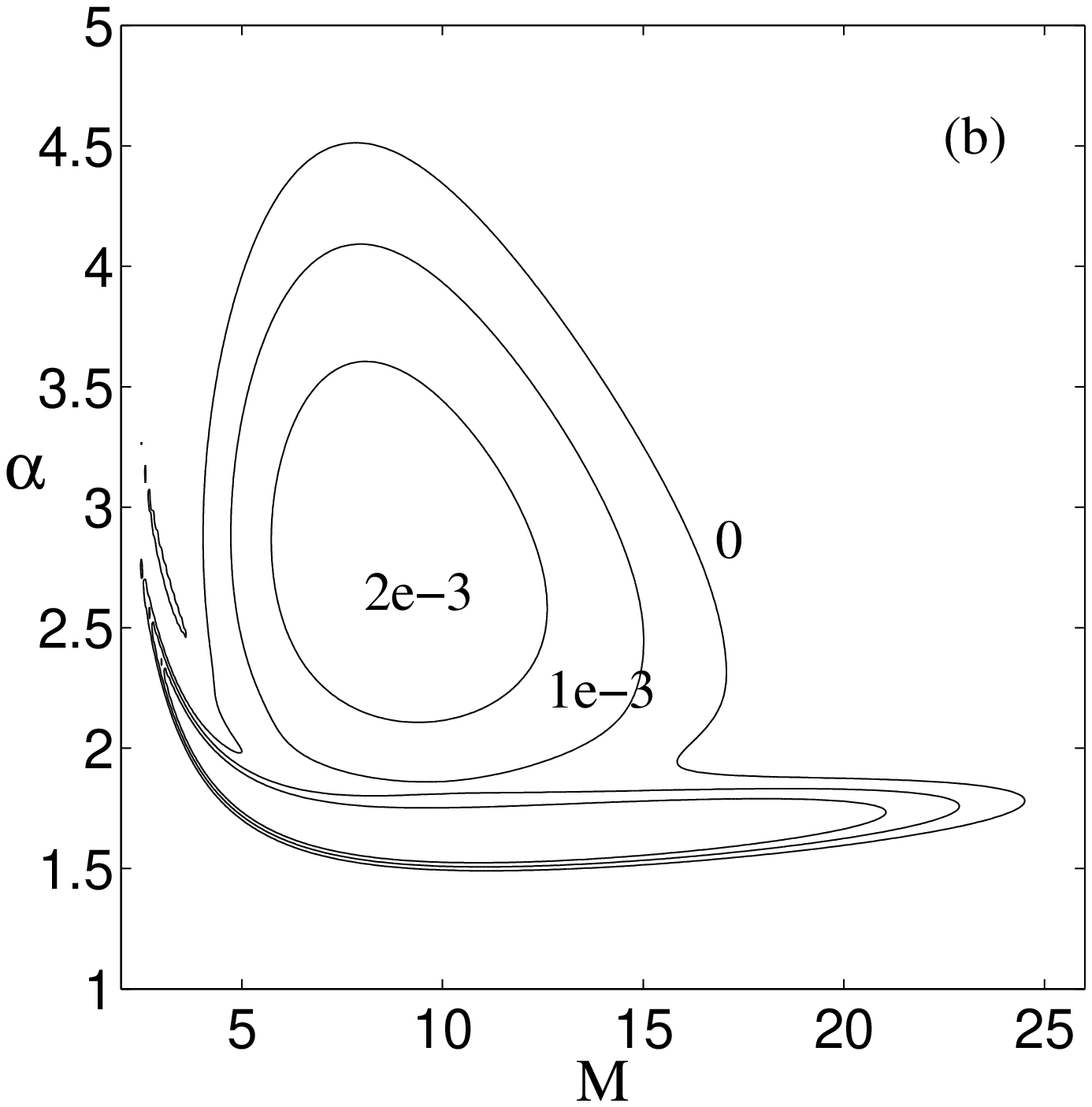}\\
\includegraphics[width=4.1cm]{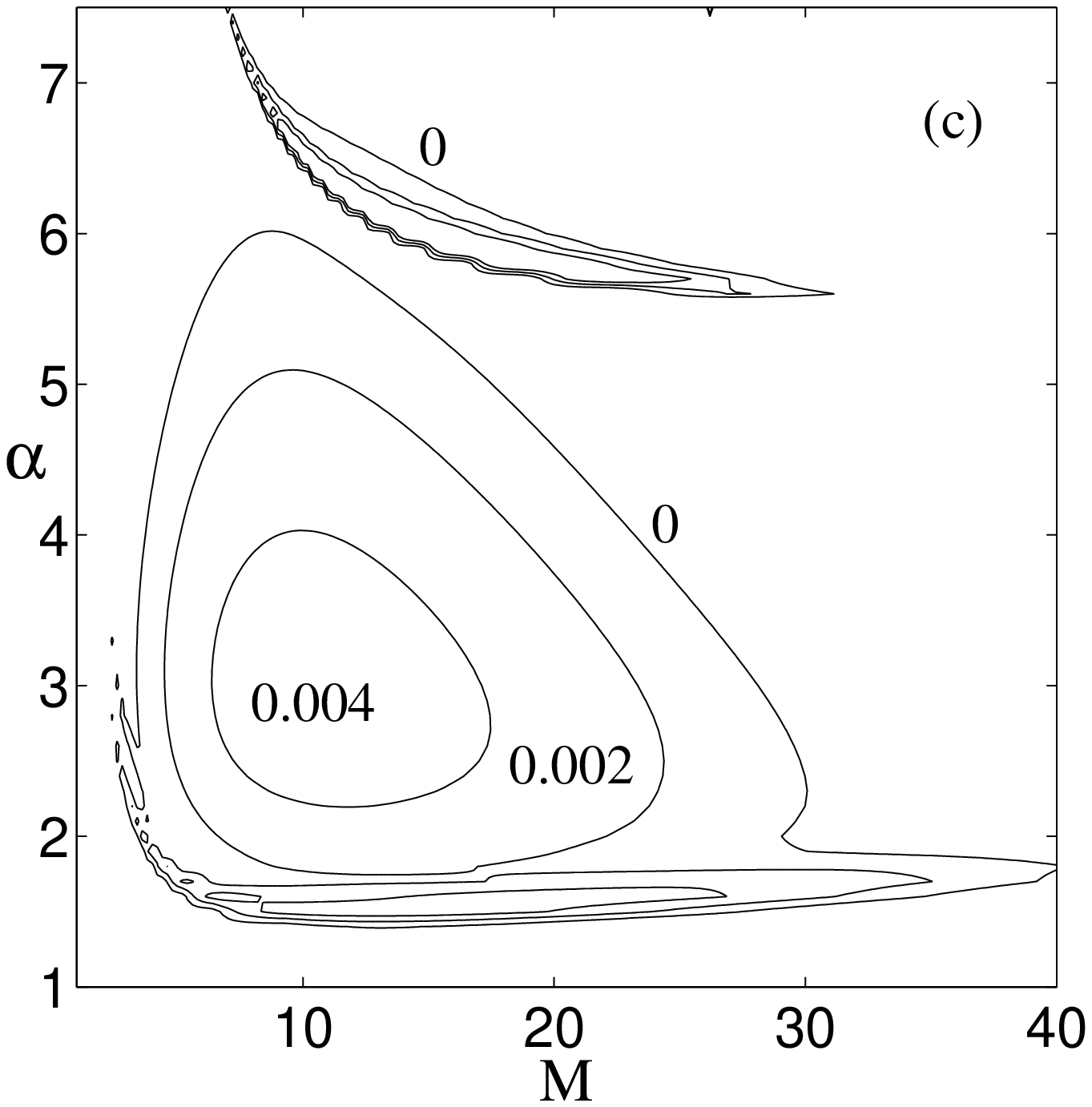}\;\;
\includegraphics[width=4.1cm]{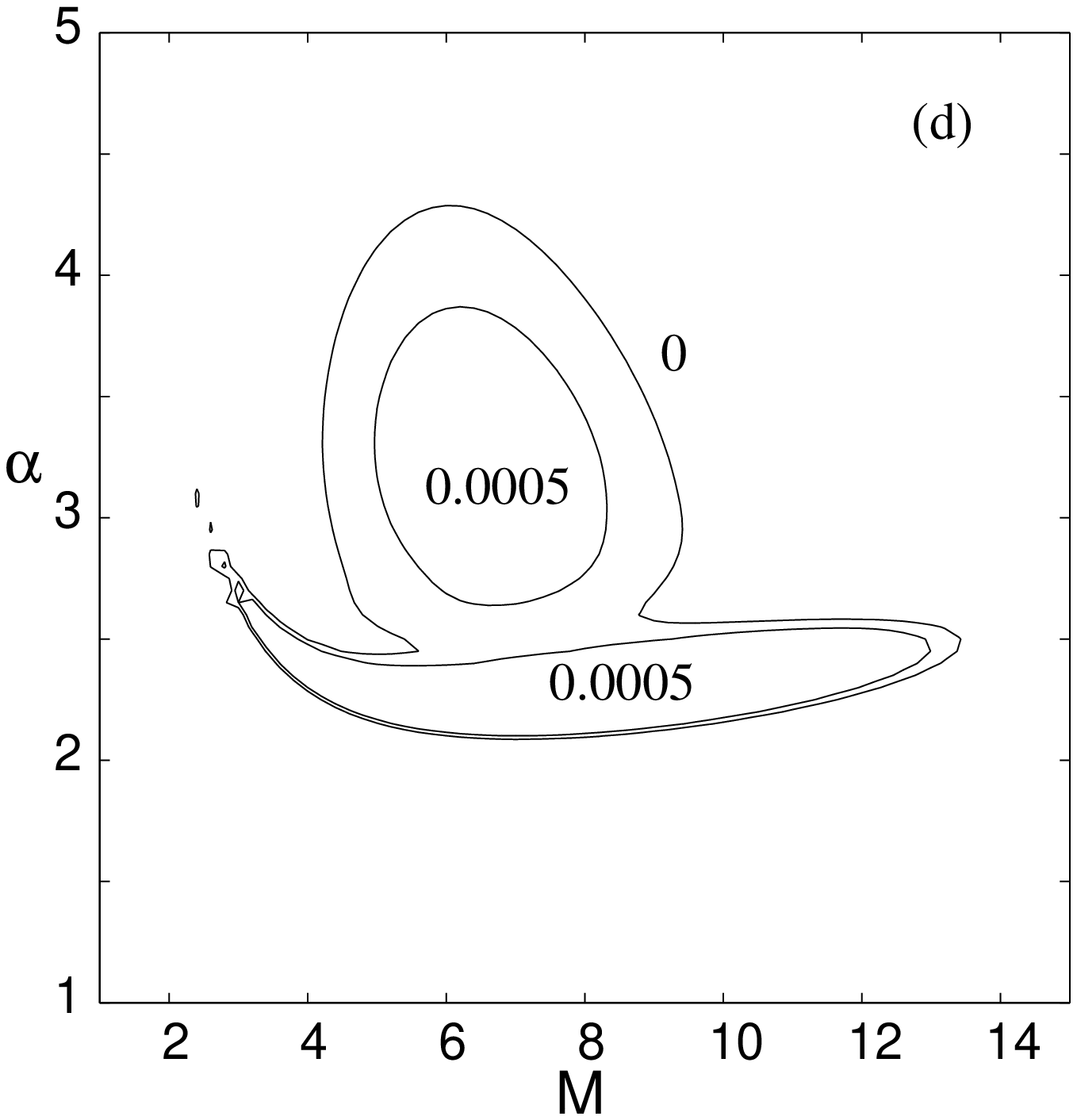}
\caption{
(a-c)  Stability maps for the uniform shear flow in the ($M, \alpha$)-plane
for two-dimensional ($\beta=0$) perturbations at different
Reynolds numbers: (a) $Re=10^5$; (b) $Re=2\times 10^5$; (c) $Re=5\times 10^5$.
Panel (d) is the analogue of panel (c) for the non-uniform shear flow
at $Re=5\times 10^5$.
In each panel, the neutral contours ($\omega_i=0$) along with a few positive growth rate
($\omega_i > 0$) contours are shown.
}
\label{fig:fig2}
\end{figure}

Figures 2(a-c) show the contours of the growth rate
of the least decaying mode, $\omega_{ldi}=\max (\omega_i)$,
in the $({\it M},\alpha)$-plane for the uniform shear flow with 
two-dimensional disturbances ($\beta=0$) at three different Reynolds numbers.
The flow is unstable inside the neutral stability contour ($\omega_{ldi}=0$) and stable outside.
With increasing $Re$, the size of the instability region increases
and there is an {\it additional instability loop} in Fig.~2(c) for $Re=5\times 10^5$.
For a comparison, the analogue of Fig.~2(c) is displayed in Fig.~2($d$)
for the non-uniform shear flow.
It is seen that the ranges of $M$ and $\alpha$, 
over which the flow is unstable, are much larger for the uniform shear flow.
Moreover, the {\it additional unstable loop} at large $\alpha$ in Fig.~2(c)
is missing in the stability diagram of the non-uniform shear flow in Fig.~2(d).
Comparing the contours of positive growth-rates in Figs.~2(c) and 2(d), we find that
the maximum growth-rate in the uniform shear flow can be 
larger by an order-of-magnitude.

\begin{figure}[h!]
\includegraphics[width=7.0cm]{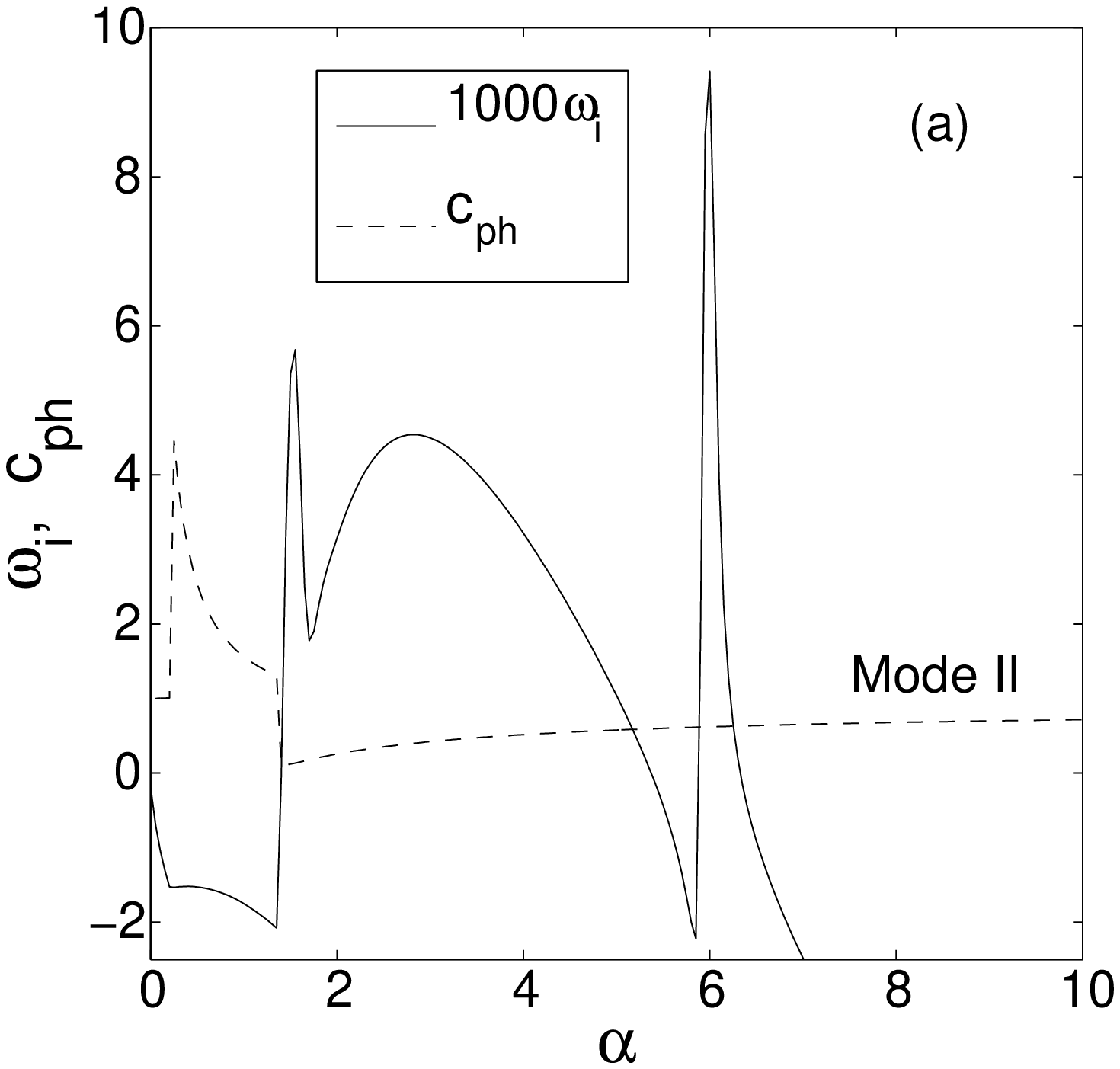}\\
\includegraphics[width=7.0cm]{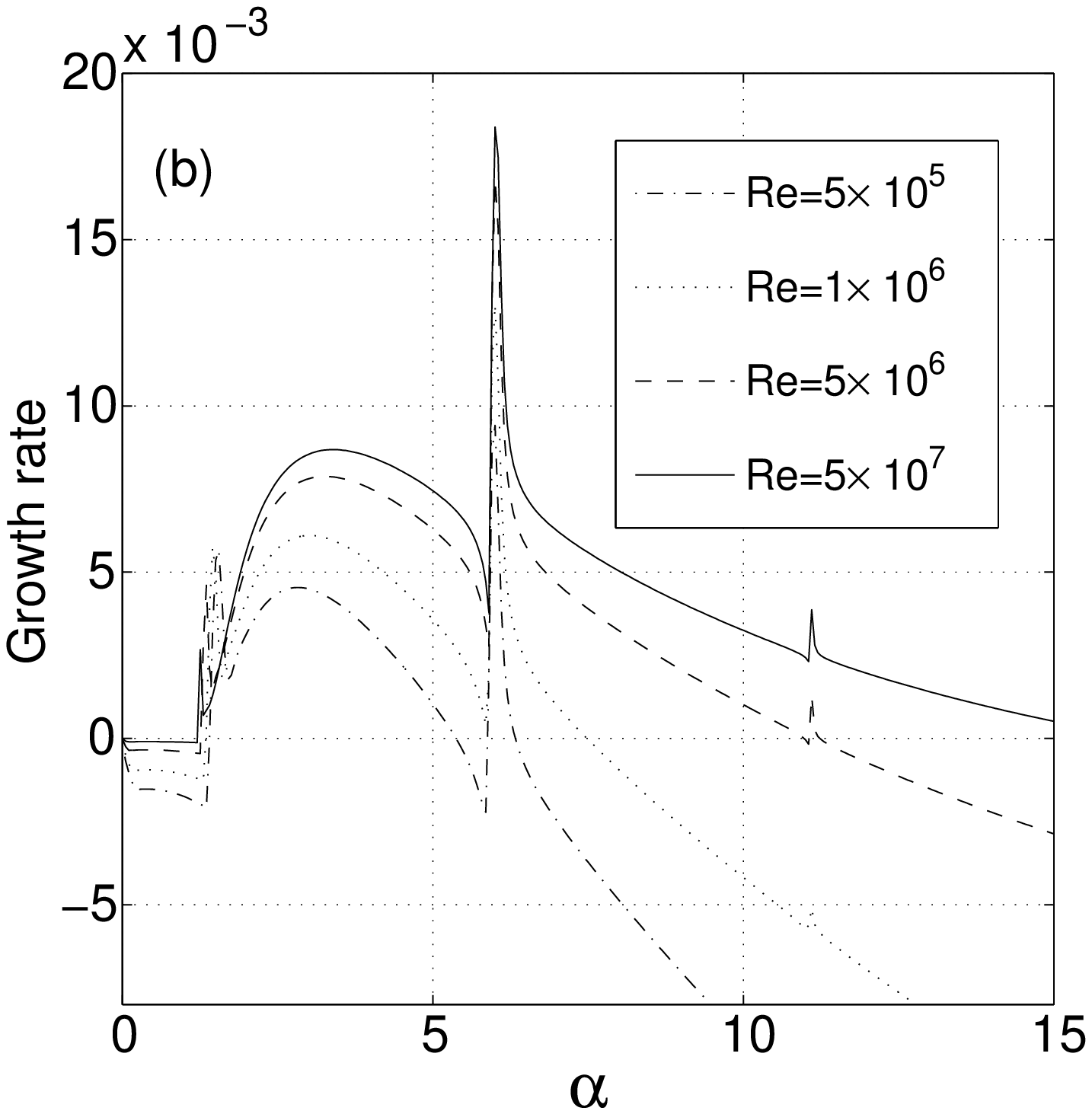}
\caption{
(a) Variations of the growth rate ($\omega_i$)
and the phase speed ($c_{ph}$) of the most unstable mode with $\alpha$ for  
two-dimensional perturbations ($\beta=0$) and $M=15$;
other parameters as in Fig.~2$c$.
(b) Effects of Reynolds number, $Re$, on
the growth rate of the most unstable mode with $\alpha$ for $\beta=0$ and $M=15$.
}
\label{fig:fig3}
\end{figure}

Figure 3(a) shows the variation of the
most unstable mode with $\alpha$ at a Mach number of $M=15$,
with other parameters as in Fig.~2(c).
The solid line denotes the growth rate ($\omega_i$) and
the dashed line the phase speed ($c_{ph}\equiv c_r=\omega_r/\alpha$).
It is observed that the flow is stable for low $\alpha$, 
but becomes unstable at $\alpha\approx 1.65$,
with the corresponding phase speed crossing {\it zero}
which implies that this instability belongs to the {\it mode-II} [see Fig.~1(b)];
the flow becomes stable again for large enough $\alpha$ ($>6.2$).
(Below $\alpha<1.65$, the mode-I is the least-stable mode
for which $c_{ph}>1$, and hence the phase-speed changes
abruptly at  $\alpha\approx 1.65$.)
Three peaks on the growth-rate curve in Fig.~3(a)
correspond to three distinct instability loops in Fig.~2(c).
It is observed that the phase speed changes 
smoothly across the kinks on the growth-rate curve for $\alpha>1.65$,
implying that there is no ``mode-crossing'' across these apparent kinks.
Hence, all three unstable peaks belong to the same mode (see following paragraph),
and, according to the above mode-classification,
the origin of this instability is  mode-II.
The effect of Reynolds numbers on the most unstable mode
is shown in Fig.~3(b), with parameter values as in Fig.~3(a).
It is observed that increasing the value of $Re$ from
$5\times 10^5$ to $5\times 10^6$, an additional
unstable peak appears on the growth-rate curve near $\alpha=11$;
however, the {\it dominant} instability (i.e., the mode having
the maximum growth-rate for all $\alpha$ for given $Re$ and $M$) still comes from the
{\it third peak} [that corresponds to the uppermost instability-lobe in Fig.~2(c)],
and this observation holds even at larger values of $Re=5\times 10^7$.

\begin{figure}[h!]
\includegraphics[width=6.8cm]{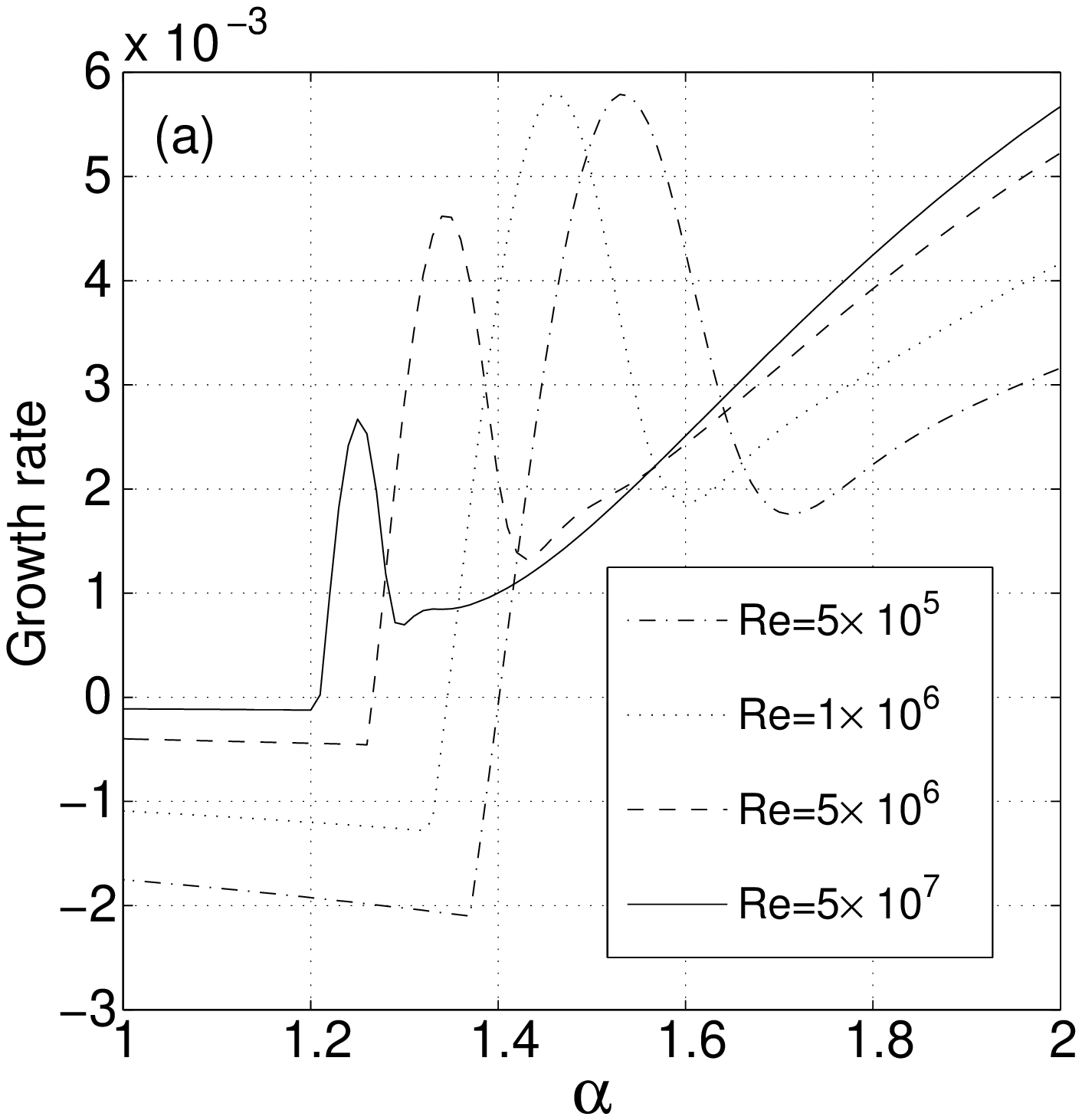}
\includegraphics[width=6.8cm]{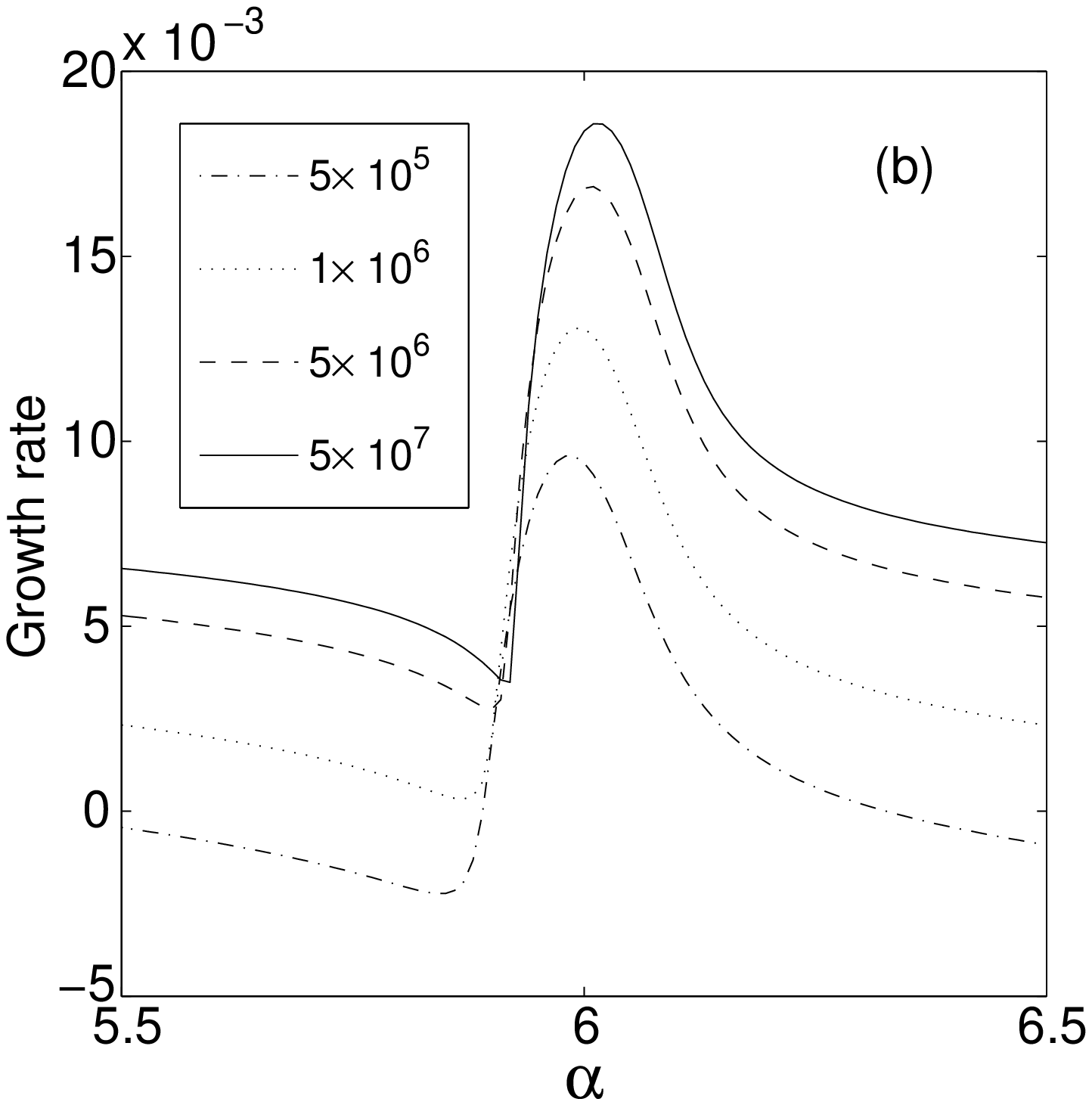}
\includegraphics[width=6.8cm]{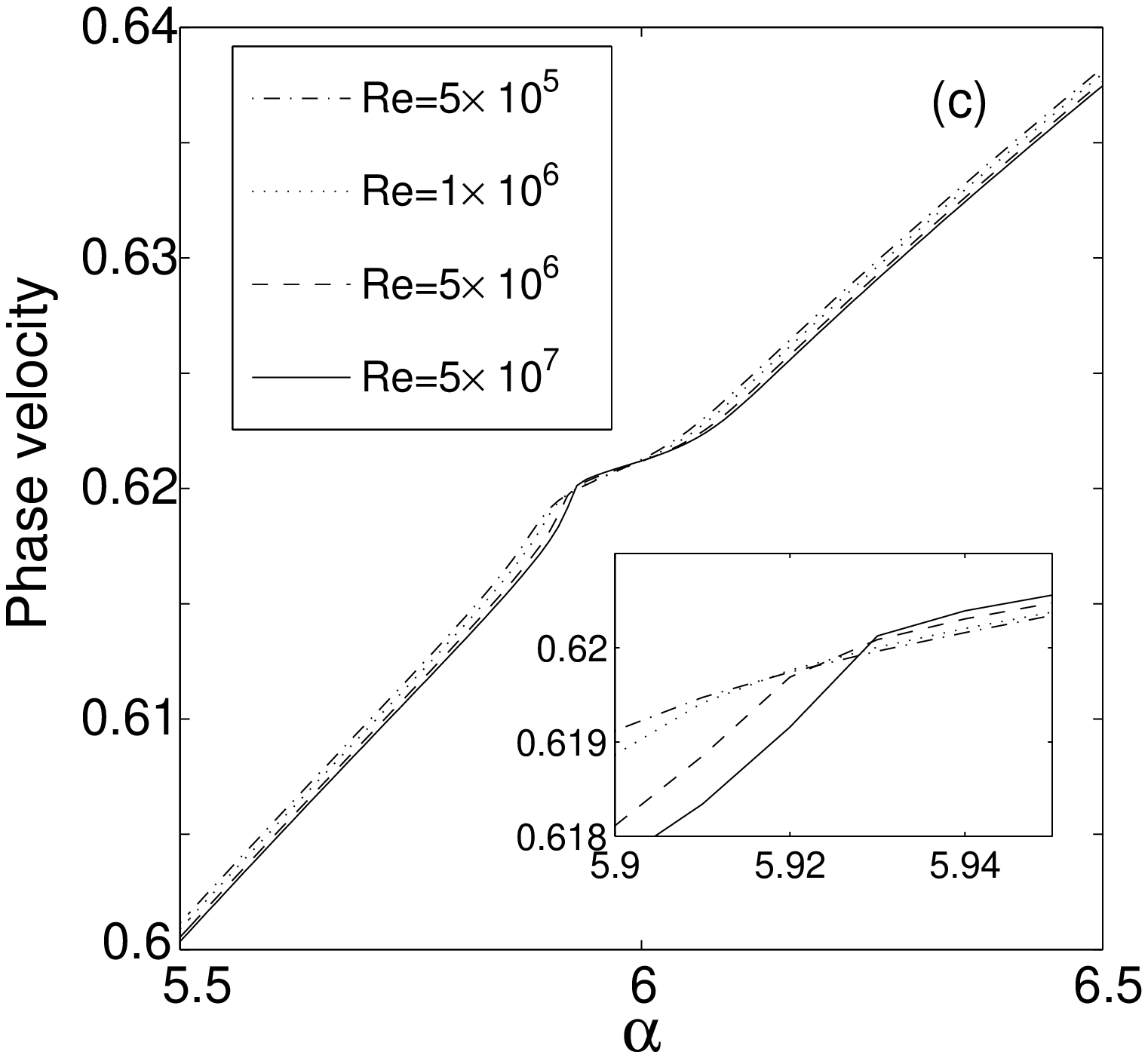}
\caption{
(a) Enlarged view of the first peak of Fig.~3$b$ around $\alpha\sim 1.5$.
(b) Enlarged view of the third peak of Fig.~3$b$ around $\alpha\sim 6$.
(c) Variation of the phase-speed curve corresponding
to the growth-rates in  panel (b).
}
\label{fig:fig4}
\end{figure}

To find out whether the sharp peaks on the growth rate curves in Fig.~3
are bounded, we show the enlarged views
of the first and third peaks [of Fig.~3(b)] in Figs.~4(a) and 4(b), respectively.
It is clear that the growth rate varies {\it smoothly} across each peak,
and the maximum growth rate at each peak is bounded; 
the {\it sharpness} of the first and third peaks in Fig.~3 is a consequence
of large variation in growth-rate (albeit smoothly) over a small range of $\alpha$.
Figure 4(c) shows the phase-speed variation corresponding
to the third-peak [i.e., Fig.~4(b)];
clearly, there is no discontinuity on the phase-speed curve too.
[The phase-speed variation across the first peak in Fig.~4(a) is 
also smooth, not shown.]
These results suggest that the instability in Fig.~3 belongs to
the same mode and the maximum growth-rate at each peak remains bounded.

From the zoom of the first peak, as shown in Fig.~4(a),
we observe that the peak-height diminishes with increasing $Re$-- 
this is a {\it viscous instability} since it disappears in the inviscid limit.
On the other hand, the height of the second, flatter, peak in Fig.~3(a)
increases with increasing $Re$ that eventually approaches the
asymptotic results on the {\it inviscid} mode II instability of Duck {\it et al.}~\cite{DEH94}.
The effect of $Re$ on the third peak in Fig.~3(a)
can be ascertained from  its enlarged version in Fig.~4(b).
This instability becomes stronger with increasing $Re$,
implying that this is an {\it inviscid} instability too.
It may be noted that this inviscid instability  was
not reported in Ref.~\cite{HZ98} for the nonuniform shear flow.

\begin{figure}[h!]
\quad\includegraphics[width=7.0cm]{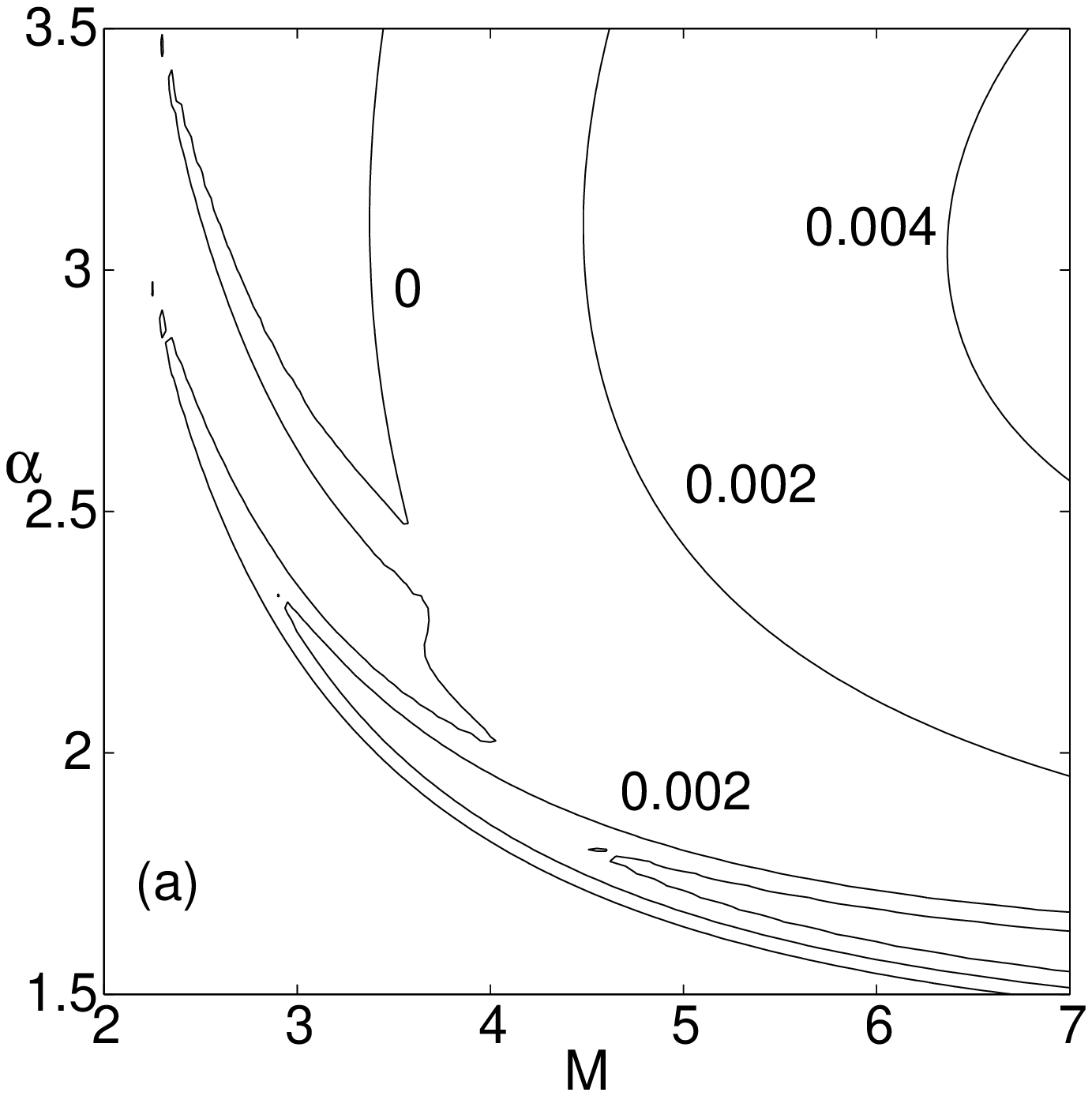}\\
\includegraphics[width=7.5cm]{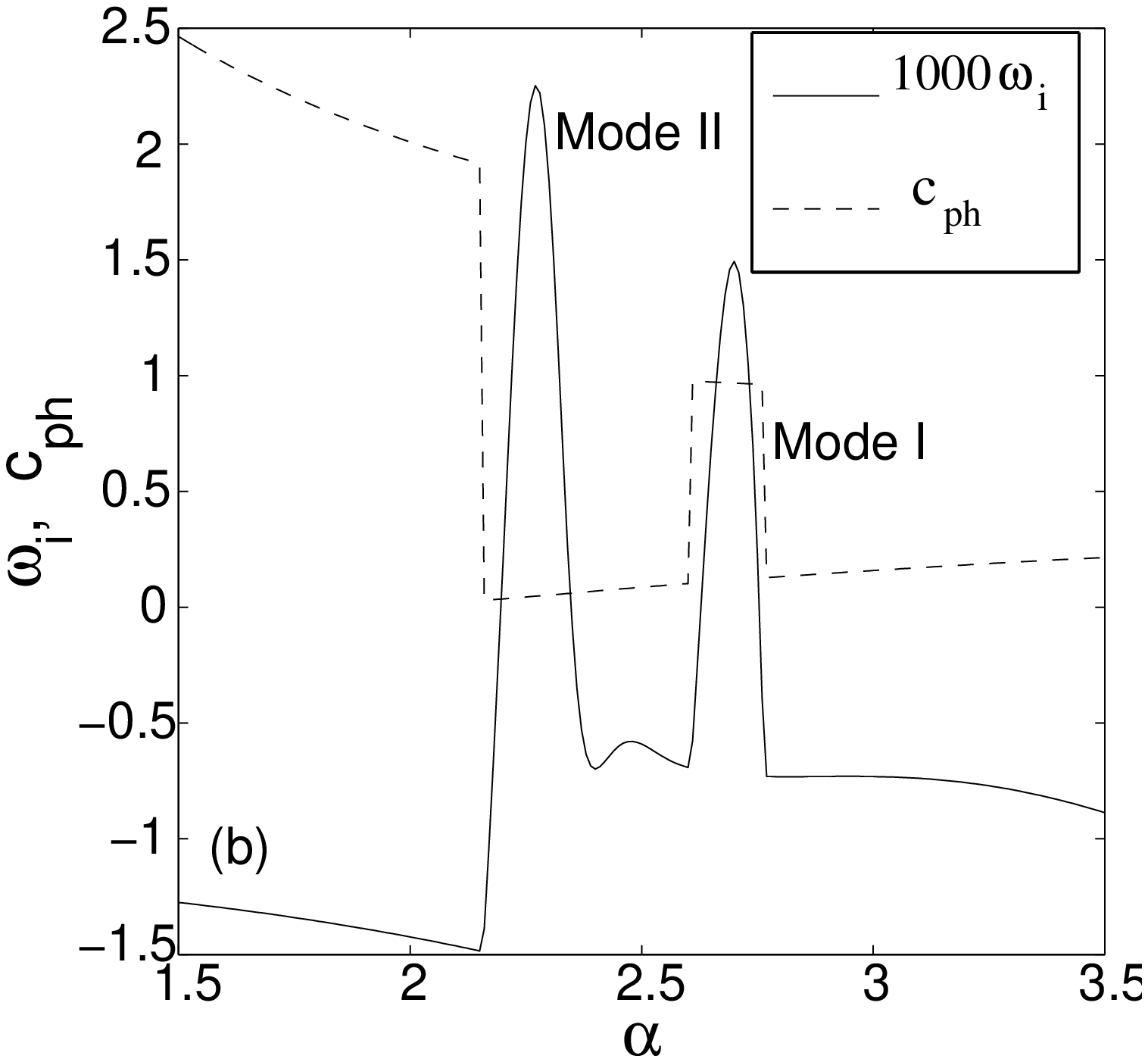}
\caption{
(a) An expanded view of the stability map in Fig.~2(c) for low Mach numbers.
(b) Variations of the most unstable mode with $\alpha$ at $M=3$.
}
\label{fig:fig5}
\end{figure}

Figure 5(a) shows the zoom of the
left hand corner of the stability map in Fig.~2(c).
There are two narrow loops of instability at $M<4$.
To find out the modal-origin of these two instability loops, we plot, in Fig.~5(b),
the variations of the growth-rate (solid line) 
and the phase speed (dashed line) of the least-stable mode with $\alpha$ at $M=3$. 
From the variation of the phase speed $c_{ph}$,
we find that the first unstable peak  is due to the mode II (phase speed near zero) and
the second peak due to the mode I (phase speed near unity).
Therefore, the upper ``narrow'' instability loop in Fig.~5(a)
belongs to mode I and the lower loop to mode II.

\begin{figure}[h!]
\includegraphics[width=7.0cm]{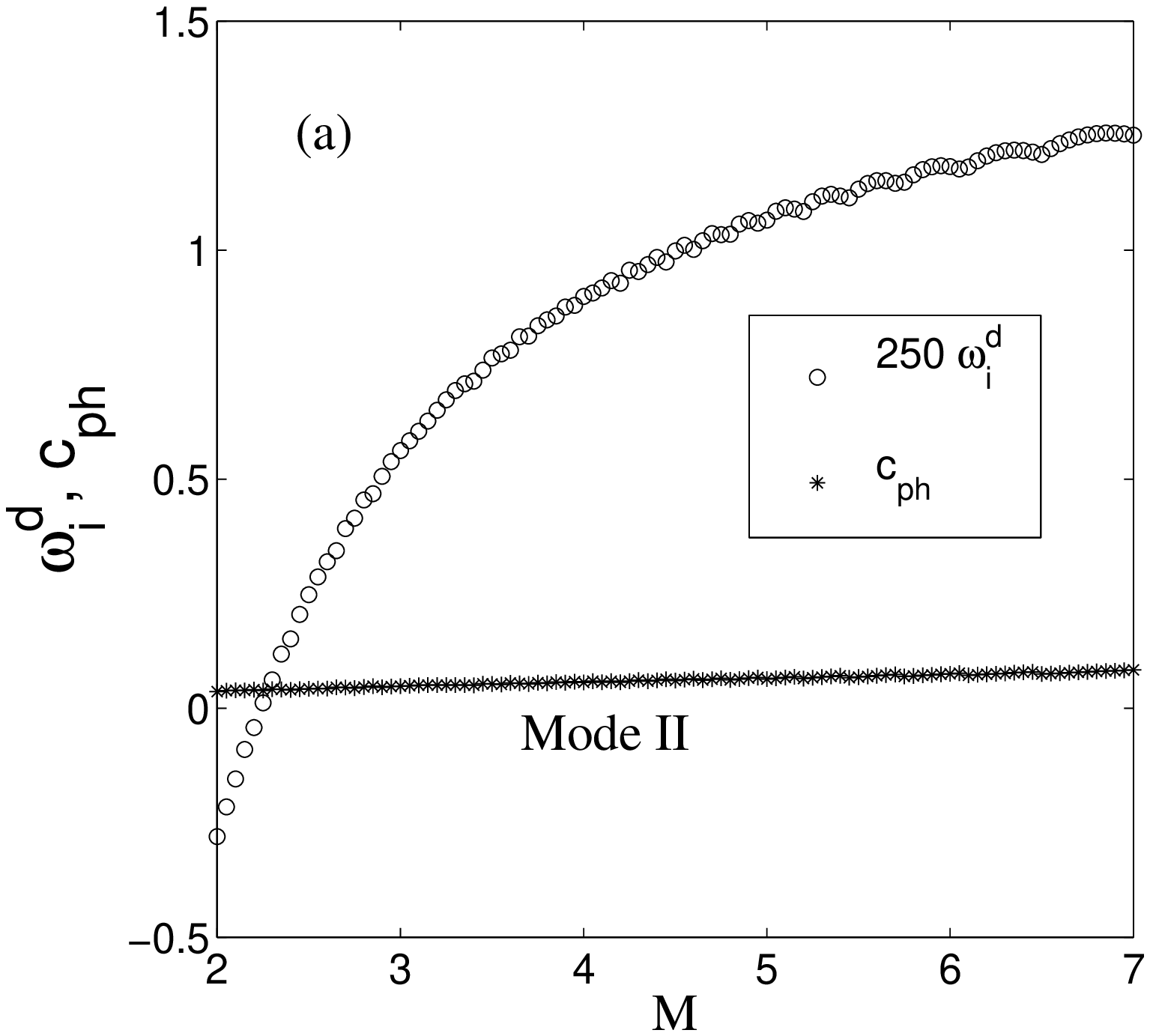}\\
\includegraphics[width=7.0cm]{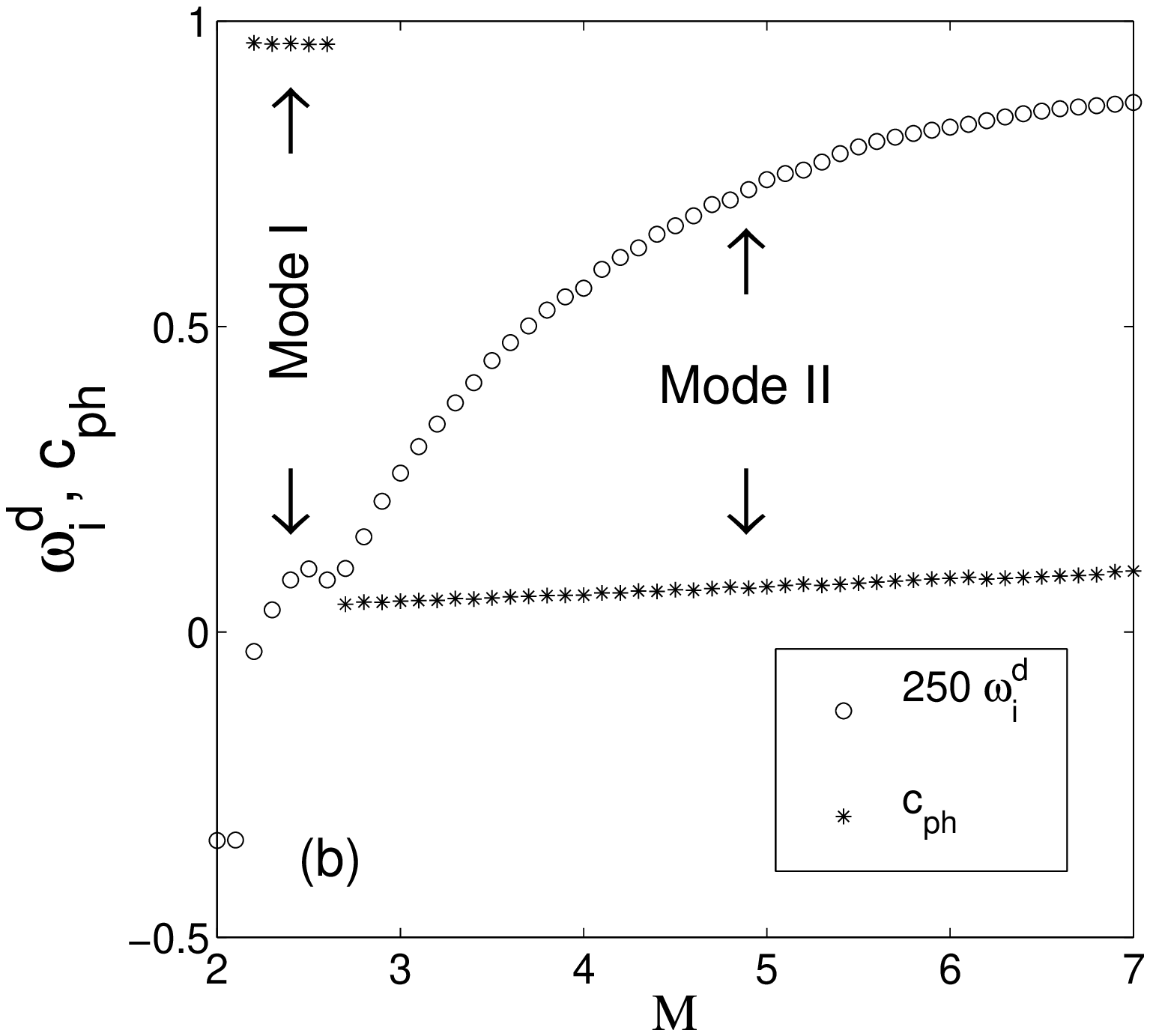}
\caption{
Variations of the maximum growth rate over all $\alpha$, $\omega_i^{d}$, 
and the corresponding phase speed, $c_{ph}$, with $M$ at $Re=5\times 10^5$:
(a) uniform shear; (b) non-uniform shear.
The arrows in panel (b) are used to identify the portions of $\omega_i^{d}$ and $c_{ph}$
over $M$ for both mode-I and mode-II.
}
\label{fig:fig6}
\end{figure}

To find out the {\it dominant} instability mode over all $\alpha$
in Fig.~5(a), we plot the variation of the maximum growth rate
\begin{equation}
    \omega_i^{d} = \max_{\alpha} \omega_i,
\label{eqn_domEig}
\end{equation}
with $M$ in Fig.~6(a) which increases monotonically with increasing $M$
for the range of $M$ shown (in fact, $ \omega_i^{d}$ decreases
beyond a critical value of $M$, see Fig.~2).
It is clear from the phase-speed variation in Fig.~6(a)
that there is no ``mode-crossing'' and 
the mode II remains the dominant instability for all $M$.
This conclusion is in contrast to the result of Hu and Zhong~\cite{HZ98}
(for non-uniform shear flow)
who found that the mode I remains the dominant mode at {\it small} $M$
and the mode II at {\it moderate-to-large} $M$, as it is evident from Fig.~6(b).
For the non-uniform shear flow,
the range of $M$ over which the mode I remains the dominant mode
increases marginally with Reynolds number (not shown for brevity). 
For example, at $Re=5\times 10^7$, the mode I is the dominant mode
for $M\sim(1.5-3)$ and the mode II for $M>3$.

\begin{figure}[h!]
\includegraphics[width=7.0cm]{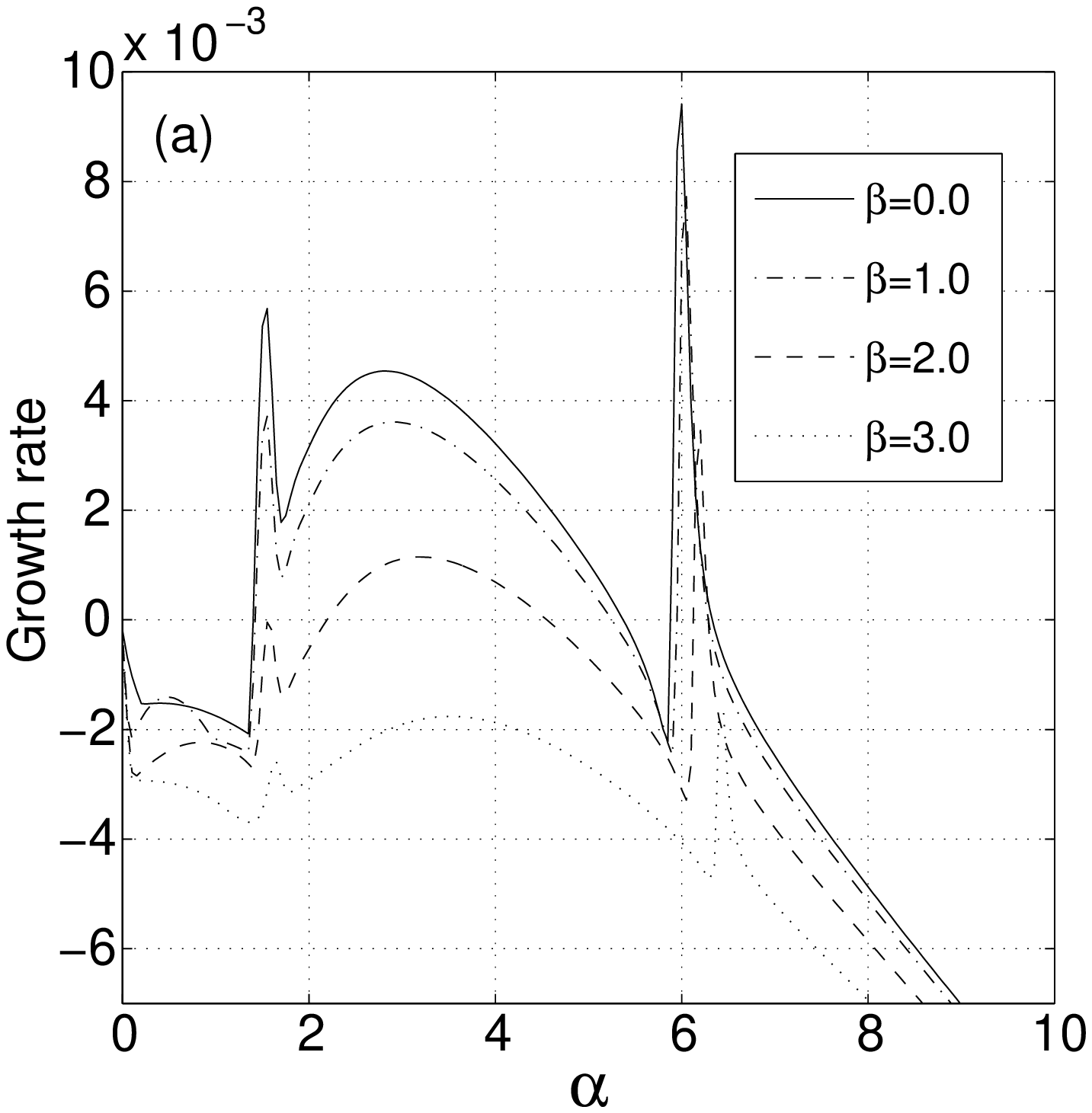}\\
\includegraphics[width=7.0cm]{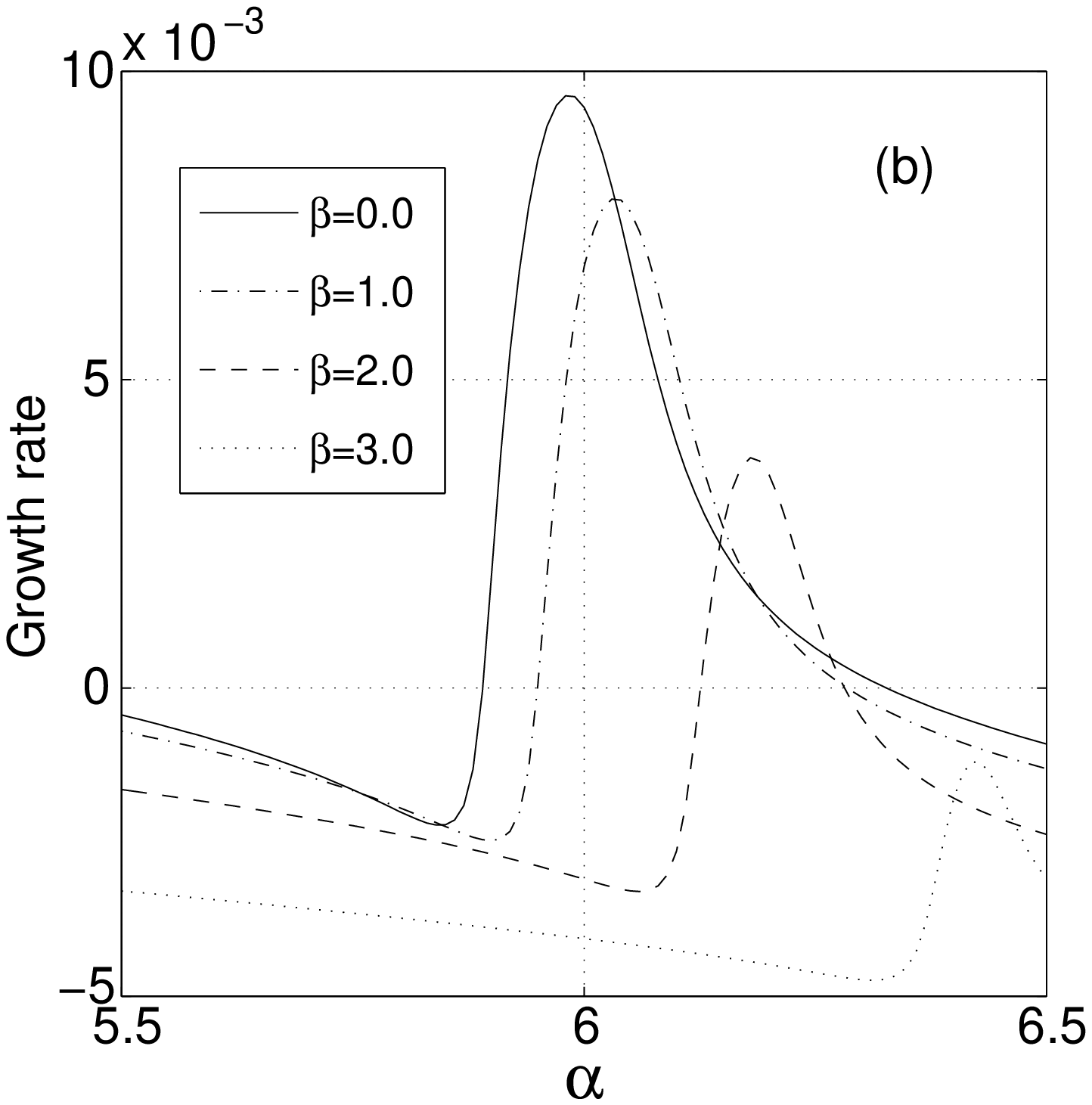}
\caption{
(a) Variations of the growth rate ($\omega_i$) of the most unstable mode
with $\alpha$ for various values of the spanwise wavenumber $\beta$ with 
$Re=5\times 10^5$ and $M=15$.
(b) Zoom of panel (a) around the third peak.
}
\label{fig:fig7}
\end{figure}

The effect of {\it three-dimensional} perturbations
on the least stable growth rate is shown  in Fig.~7(a)
for different span-wise wavenumber $\beta$,
with parameter values $Re=5\times 10^5$ and $M=15$;
the zoom of the third-peak is displayed in Fig.~7(b).
Comparing different growth-rate curves with the one for two-dimensional 
perturbations ($\beta=0$), we find that there is a window of 
$\alpha$, slightly beyond the third-peak, over  which
the three-dimensional perturbations are more unstable 
than their two-dimensional counterparts.
Therefore, in general, Squire's theorem is not valid for the present flow configuration.
This finding  is in variance with the previous  work~\cite{Glat89}
that Squire's theorem holds irrespective of the value of $\alpha$
in the uniform shear flow of an ``isothermal'' compressible fluid.

\subsection{{\label{CritReynolds}}Critical Reynolds Number}

\begin{figure}[h!]
\includegraphics[width=7.2cm]{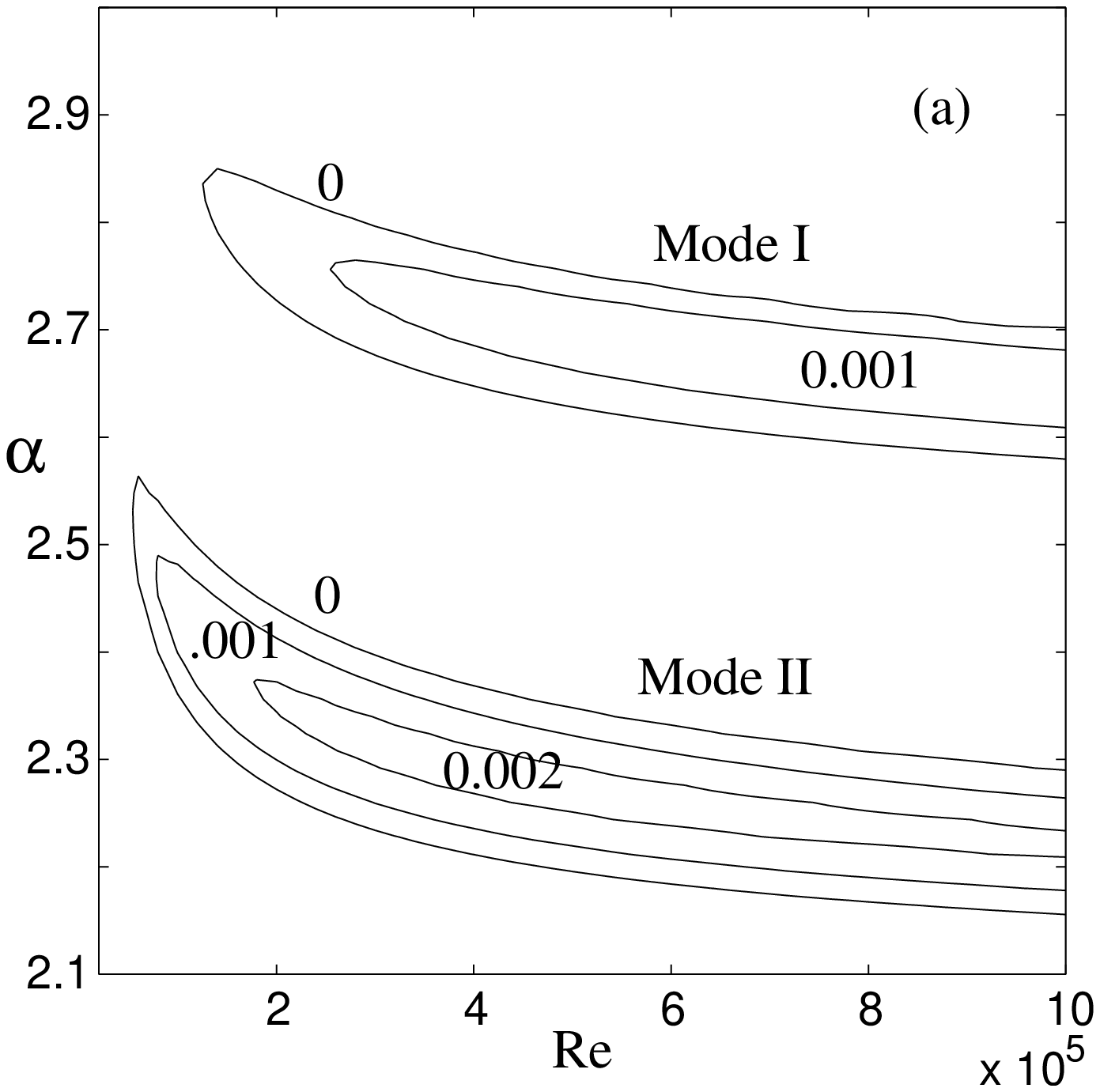}\\
\includegraphics[width=7.0cm]{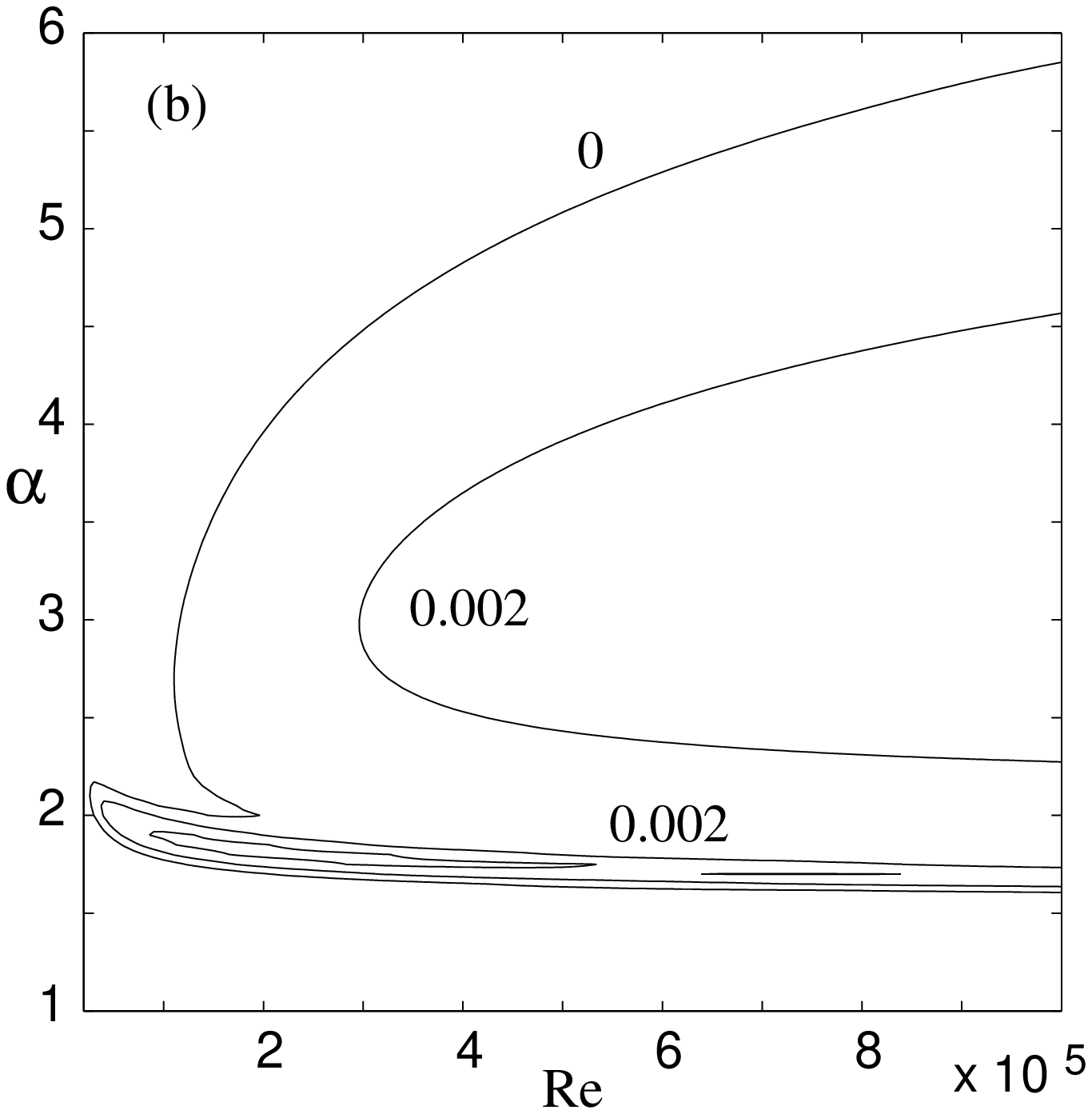}
\caption{
Stability maps for uniform shear flow in the ($Re, \alpha$)-plane
for two-dimensional ($\beta=0$) perturbations: (a) $M=3$; (b) $M=5$.
In each panel, the neutral contours ($\omega_i=0$) along with a few positive 
growth rate ($\omega_i>0$) contours are shown.
}
\label{fig:fig8}
\end{figure}

Figures 8(a) and 8(b) show the contours of the 
least stable growth rate in the $({\it Re},\alpha)$-plane for 
two-dimensional disturbances ($\beta=0$) with $M=3$ and $M=5$, respectively.
The upper and lower instability loops in Fig.~8(a)
correspond to mode I and mode II instability, respectively,
whereas the instability loop in Fig.~8(b) arises solely from mode II.
For  $M=3$, the flow becomes unstable to {\it mode I} at
$(Re, \alpha)\approx (123900,2.835)$,
and to {\it mode II} at $(Re, \alpha)\approx (50060,2.545)$.
Therefore, the critical Reynolds number ($Re_{cr}$) at which
the instability sets in {\it first} is determined by  {\it mode II} 
in uniform shear flow --
this observation holds at other values of $M$.
A comparison of the values of $Re_{cr}$ and $\alpha_{cr}$
between the uniform and non-uniform shear flows is given in Table I
for different  Mach numbers. It is clear that 
the critical Reynolds number for the uniform shear flow is
significantly smaller than that for its non-uniform counterpart;
for example, at $M=10$, $Re_{cr}$ of two mean flows differ by a factor of $5.6$.
Therefore, we conclude that {\it the viscosity-stratification of the base-flow
would lead to a ``delayed'' transition in compressible Couette flow in terms of
modal instability}.
Another interesting observation in Table I is that
the variation of  $Re_{cr}$ with $M$ is {\it non monotonic} in the sense that 
the critical Reynolds number 
reaches a minimum at some intermediate value of Mach number.

\begin{center}
\begin{tabular}{||c|c|c|c|c||}\hline
 $ $ & \multicolumn{2}{||c|}{Uniform Shear} &\multicolumn{2}{||c||}{Non-uniform Shear} \\ \hline
  Mach Number & $Re_{cr}$ & $\alpha_{cr}$ & $Re_{cr}$ & $\alpha_{cr}$   \\ \hline
 M=3 & 50\,060 & 2.545 & 1\,64\,900 & 2.840 \\ \hline
 M=5 & 23\,830 & 2.130 & 85\,725 & 2.570 \\ \hline
 M=10 & 45\,040  &1.870  & 2\,52\,700 & 2.485 \\ \hline
 M=15 & 85\,150  &1.810  & 6\,55\,850 & 2.490 \\ \hline
\multicolumn{5}{||c||}{TABLE I: Critical stability parameters for $\beta=0$} \\ \hline
\end{tabular}
\end{center}

The effect of Reynolds number on the mode I instability
[upper loops in Fig.~5(a) and Fig.~8(a)] is to make 
it a {\it neutral} mode in the inviscid limit
as is the case for non-uniform shear flow~\cite{DEH94}.
This effect is similar to the first peak mode II instability in Fig.~4(a)
where the viscosity plays a destabilizing role.
Therefore, while the viscosity plays a {\it dual} role of destabilizing
[at small $\alpha$ as in Fig.~ 4(a)] and stabilizing
[at moderate-to-large  $\alpha$ as in Fig.~4(b)]
the mode II instability, it destabilizes the mode I instability.
This conclusion  also holds for the non-uniform shear flow~\cite{HZ98}.

Even though we have presented all stability results on mode-I and mode-II instabilities,
it may be noted out that the higher-order even (IV,...) and odd (III,...) 
inviscid modes can also become unstable
but they remain {\it sub-dominant} with respect to  mode-II instability.

\subsection{{\label{Energyanal}}Energy Analysis: Instability Mechanism}

The exponential instability can be understood by considering the rates of transfer of 
energy by the different terms in the momentum and thermal equations. 
For this we need to define a suitable norm of the perturbations which can represent the energy. 
We define the perturbation energy density as
\begin{equation}
  {\cal E}(\alpha,\beta,t) = \int_{0}^{1}{\bf \tilde{q}}^{\dagger}(y,t)
      {\mathcal M}{\bf \tilde{q}}(y,t){\rm d}y,
\label{eqn_energyden1}
\end{equation}
where the superscript $\dagger$ on any quantity refers to its conjugate value,
and the weight matrix ${\mathcal M}$ is diagonal and positive definite.
Among various choices of the weight matrix ${\mathcal M}$,
we consider the following:
\begin{equation}
 {\mathcal M} = {\rm diag}\{\rho_0, \rho_0, \rho_0, T_0/\rho_0\gamma M^2, 
     \rho_0/\gamma(\gamma-1)T_0M^2 \},
\label{eqn_WMatrix1}
\end{equation}
that corresponds to the well-known Mack-norm~\cite{Mack84} that has been used in many 
transient growth studies on compressible flows~\cite{HSH96,MAD06}. 
A special property of this norm is that this definition of energy is free 
from any contribution due to the pressure related terms in the governing equations.

Equation (\ref{eqn_energyden1}) can be written for the least decaying mode, 
which has an exponential time dependence, as
\begin{equation}
  {\cal E}_{ld}(\alpha,\beta,t) = \exp[2\Im(\omega_{ld})t]\int_{0}^{1}{\bf q'_{ld}}^{\dagger}(y)
      {\mathcal M}{\bf q'_{ld}}(y){\rm d}y,
\label{eqn_energyld}
\end{equation} 
where the subscript 'ld' refers to 'least-decaying' mode.
The rate of change of this energy with respect to time can be written as
\begin{equation}
\frac{\partial {\cal E}_{ld}}{\partial t} = 2\Im(\omega_{ld}) 
    \exp[2\Im(\omega_{ld})t]\int_{0}^{1}{\bf q'_{ld}}^{\dagger}(y)
      {\mathcal M}{\bf q'_{ld}}(y){\rm d}y,
\label{eqn_energyrateld}
\end{equation}
which can be manipulated using equation~(\ref{eigeqn}) to yield
\begin{equation}
\frac{\partial {\cal E}_{ld}}{\partial t} = 
   - {\rm i} \exp[2\Im(\omega_{ld})t]\int_{0}^{1}{\bf q'_{ld}}^{\dagger}(y)
      {\mathcal M}{\cal L}{\bf q'_{ld}}(y){\rm d}y + c.c.
\label{eqn_energyrate2ld}
\end{equation}

Now, we decompose the total energy-transfer-rate into 
those coming and going through different physical routes.
\begin{equation}
    \frac{\partial {\cal E}_{ld}}{\partial t} = 
            \exp[2\Im(\omega_{ld})t] \sum_{j = 0}^4 \dot{\cal E}_j,
\label{eqn_energyrate_decomp}
\end{equation}
where the explicit forms of the $\dot{\cal E}_j$'s are given in the appendix. 
$\dot{\cal E}_0$ is the energy-transfer-rate due to the convection by mean flow, 
$\dot{\cal E}_1$ is the same from the mean flow to the perturbation, 
$\dot{\cal E}_2$ is due to viscous dissipation, 
$\dot{\cal E}_3$ is due to the thermal diffusion, 
and finally $\dot{\cal E}_4$ is due to the viscous dissipation term 
in the thermal energy equation.

Note that the above  expressions involve the eigenfunction of the least-stable
mode and its derivative. 
The numerical estimation of these quantities is a challenging one for the least-decaying mode
at high $Re$ and $M$ with large $\alpha$ and $\beta$. 
The streamwise velocity and temperature perturbations exhibit boundarylayer like steep 
variations near the wall. These variations are extremely rapid at high $\alpha$. 
Moreover,  at high $\alpha$ there are also internal layers. 
An accurate estimation of the above quantities will require a highly resolved 
scheme to capture these steep variations. 
Therefore we used a multidomain spectral calculation, with appropriate matching conditions 
which can be found in~\cite{Malik90,MAD08} except that we have relaxed the matching of the 
derivative of the density perturbation, since the highest order of  density is one 
in the continuity equation. A check on the accuracy of the results 
has been  made by estimating 
the energy transfered by the pressure terms which must be  vanishingly small by the 
definition of the Mack energy norm.

\begin{figure}[h!]
\includegraphics[width=7.0cm]{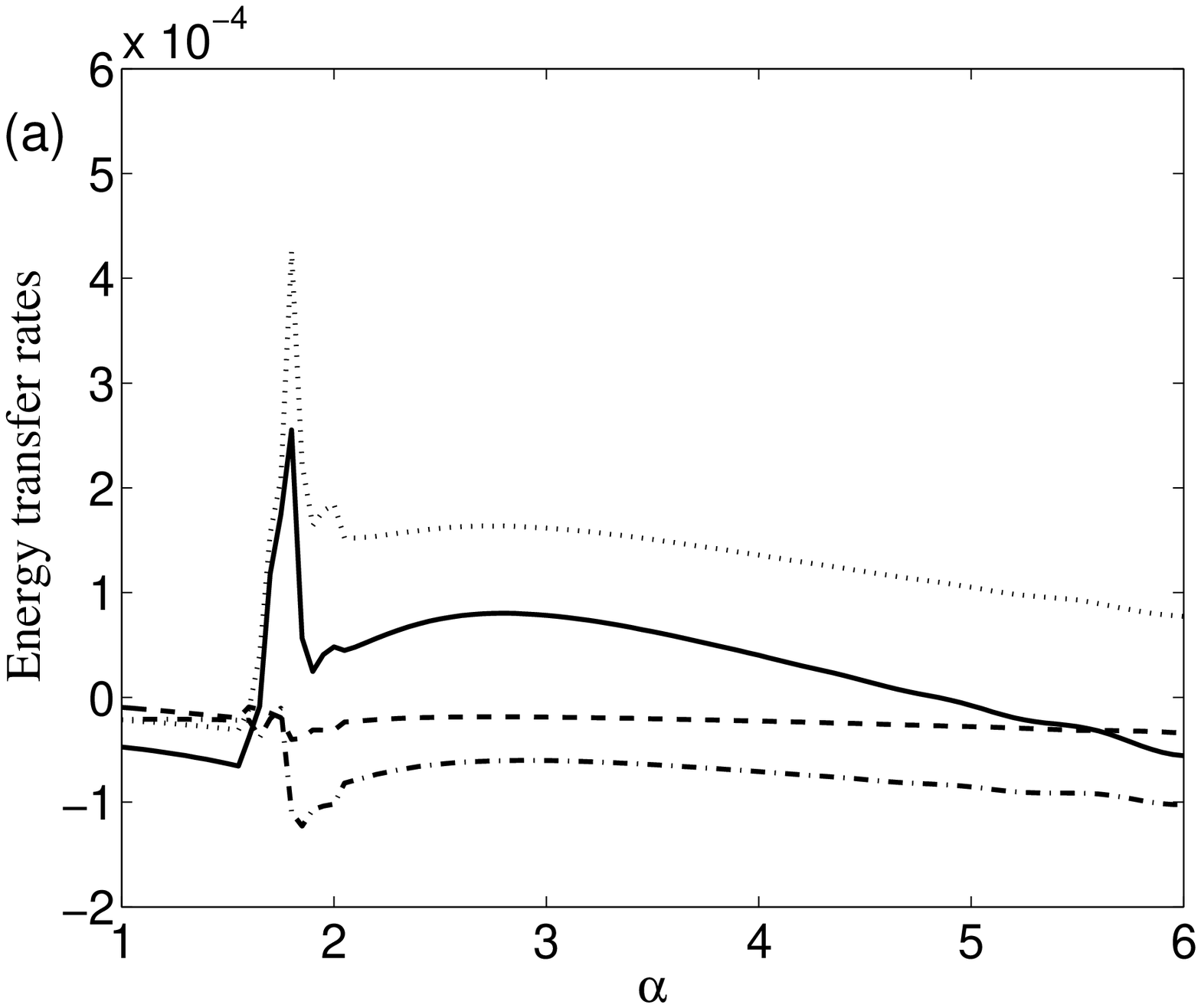}\\
\includegraphics[width=7.0cm]{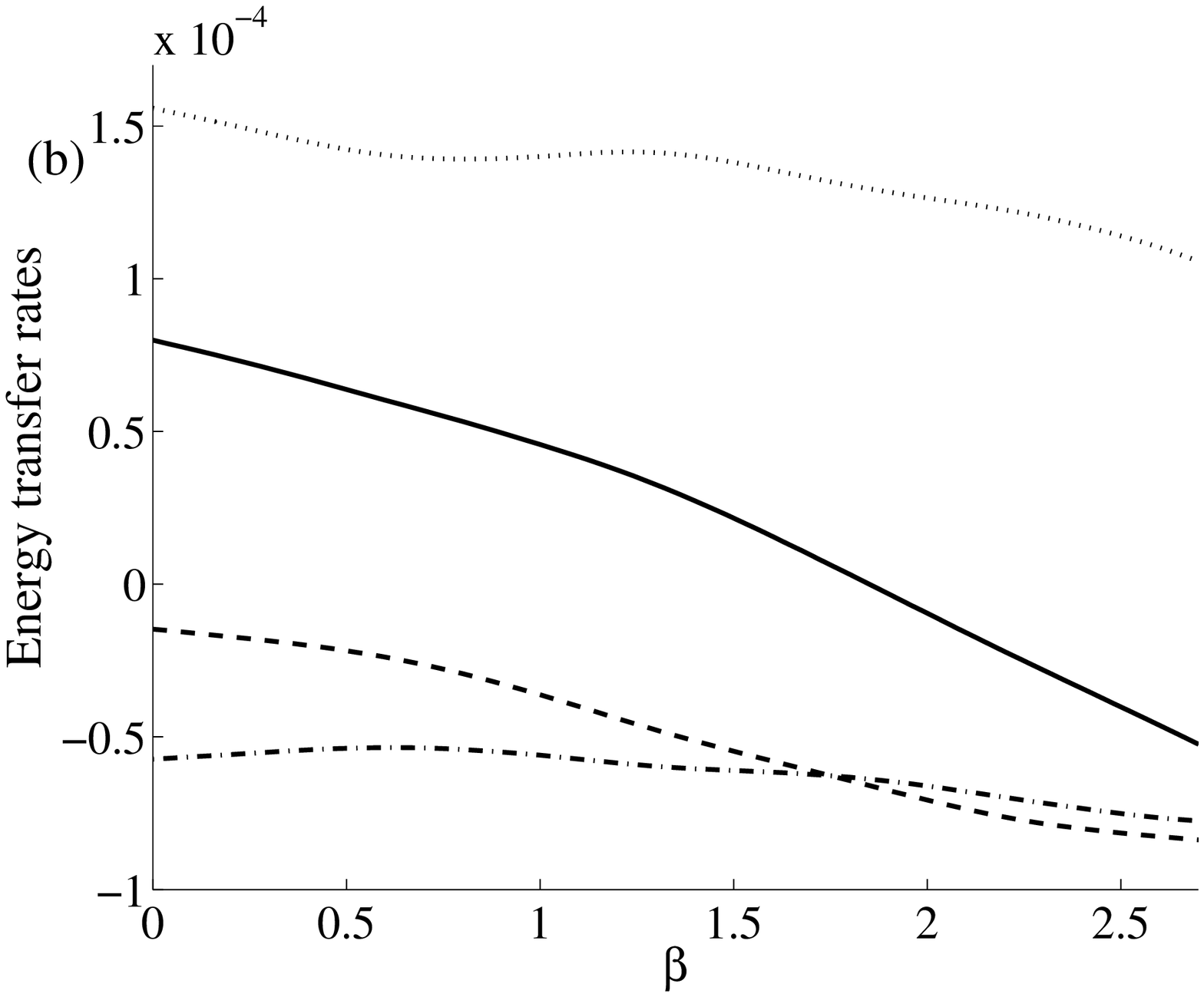}
\caption{
Rates of transfer of different energies [${\cal E}_i$, see Eq. (20)] 
for ${\it Re} = 4 \times 10^5$ at $M = 5$. 
Solid line, total energy-transfer rate; dashed line, viscous dissipation; 
dotted line, from base-flow; dash-dotted line, thermal diffusion. 
(a) $\beta = 0$; (b) $\alpha = 3$.
}
\label{fig:fig9}
\end{figure}

Figure~9 shows the rates of different constituent energies routed via
different physical processes at $M = 5$ for ${\it Re} = 4 \times 10^5$. 
In this figure, $\dot{\cal E}_4$ is not shown since it is negligibly small. 
Figure~9(a) shows results  for 2D modes for a range of $\alpha$. 
The sudden changes for $1.5 < \alpha < 2$ is due to a mode-crossing. 
The energy transfered from the mean-flow plays a dominant role for the onset of instability. 
The viscous dissipation and thermal diffusion plays the role of routing the energy 
out of perturbations; it is interesting to note that the thermal diffusion 
rate is dominant over the rate of viscous dissipation for 2D modes. 
Figure~9(b) shows these energy transfer rates for 3D modes 
for a range of $\beta$ with $\alpha = 3$.  The main difference is that at high values 
of $\beta$ the viscous dissipation dominates over thermal dissipation
for 3D modes. This observation holds at other values of $M$ and $Re$.

\begin{figure}[h!]
\includegraphics[width=7.5cm]{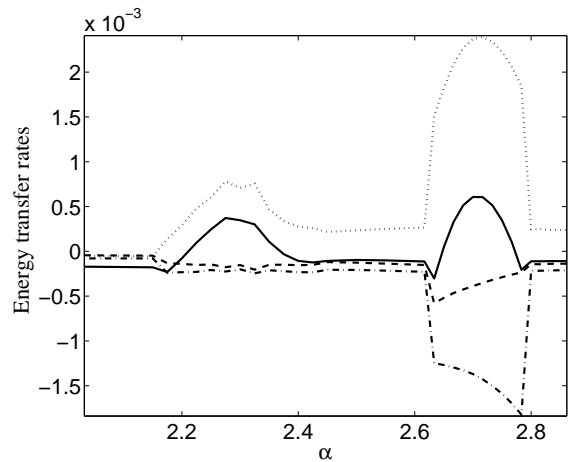}
\caption{Same as Fig.~9(a) but for M=3.}
\label{fig:fig10}
\end{figure}

As shown in Fig.~8(a) there are distinct regions of instabilities
in the (${\it Re}-\alpha$)-plane due to mode I and mode II. 
In order to study the characteristics special to each of these modes, 
we show the budget of energy-transfer-rates 
across a range of $\alpha$ spanning two different regions of instabilities
in Fig.~10, with parameter values as in Fig.~8(a). 
Both mode I and mode II instability regions exhibit a qualitatively similar behavior 
in the shares of each physical 
processes except that the balancing involved is quantitatively different for each mode.
For mode I instability, the energy transfer rate from the mean flow 
and the thermal-diffusion rate are  much larger than those for mode II.

\begin{figure}[h!]
\begin{center}
\begin{tabular}{c}
\begin{minipage}[t]{3.0in}
\begin{picture}(3.0,2.5)
\centerline{\psfig{figure=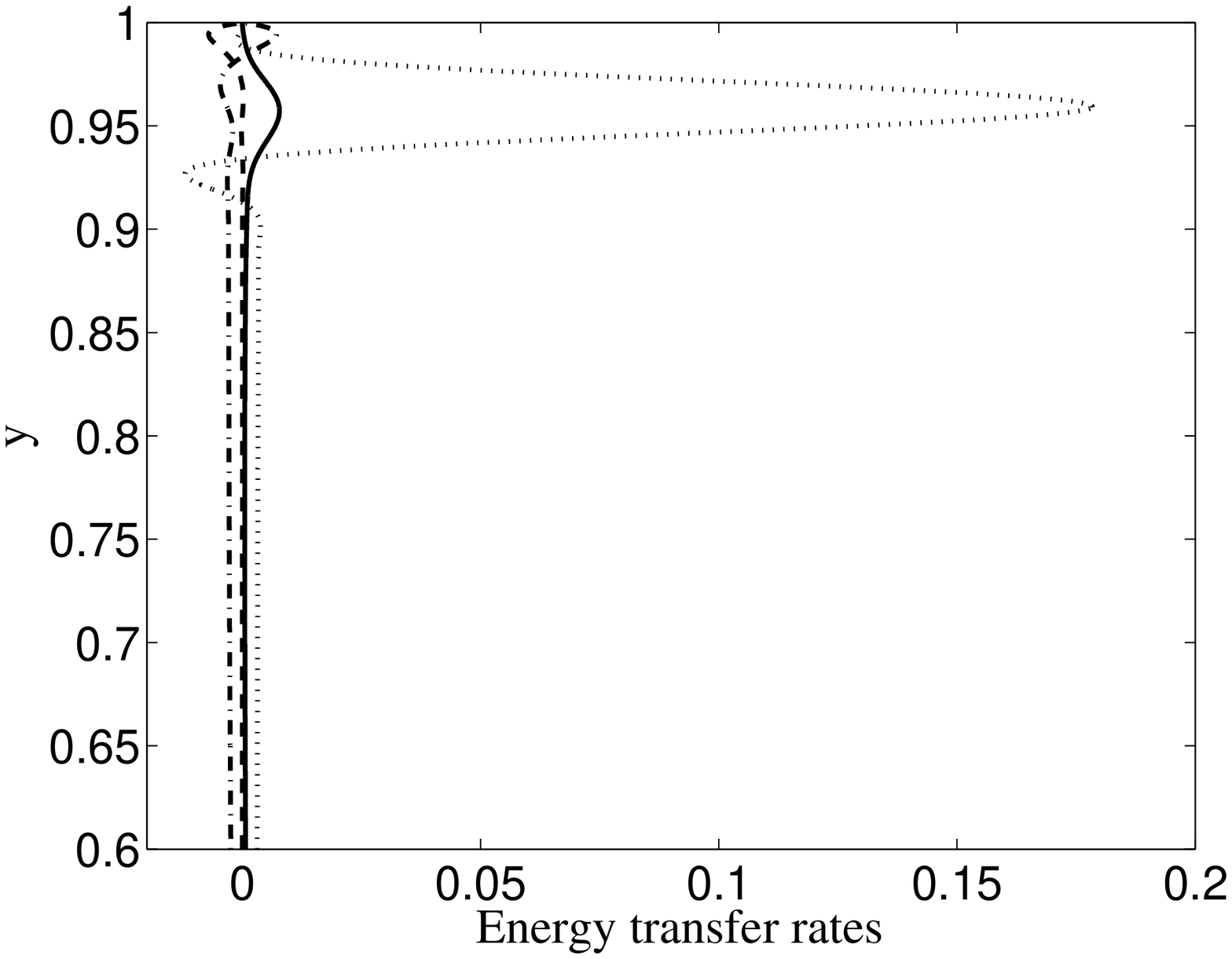 ,width=3.0in}}
\put(-3.0,2.1){(a)}
\put(-2.2,0.4){\psfig{figure=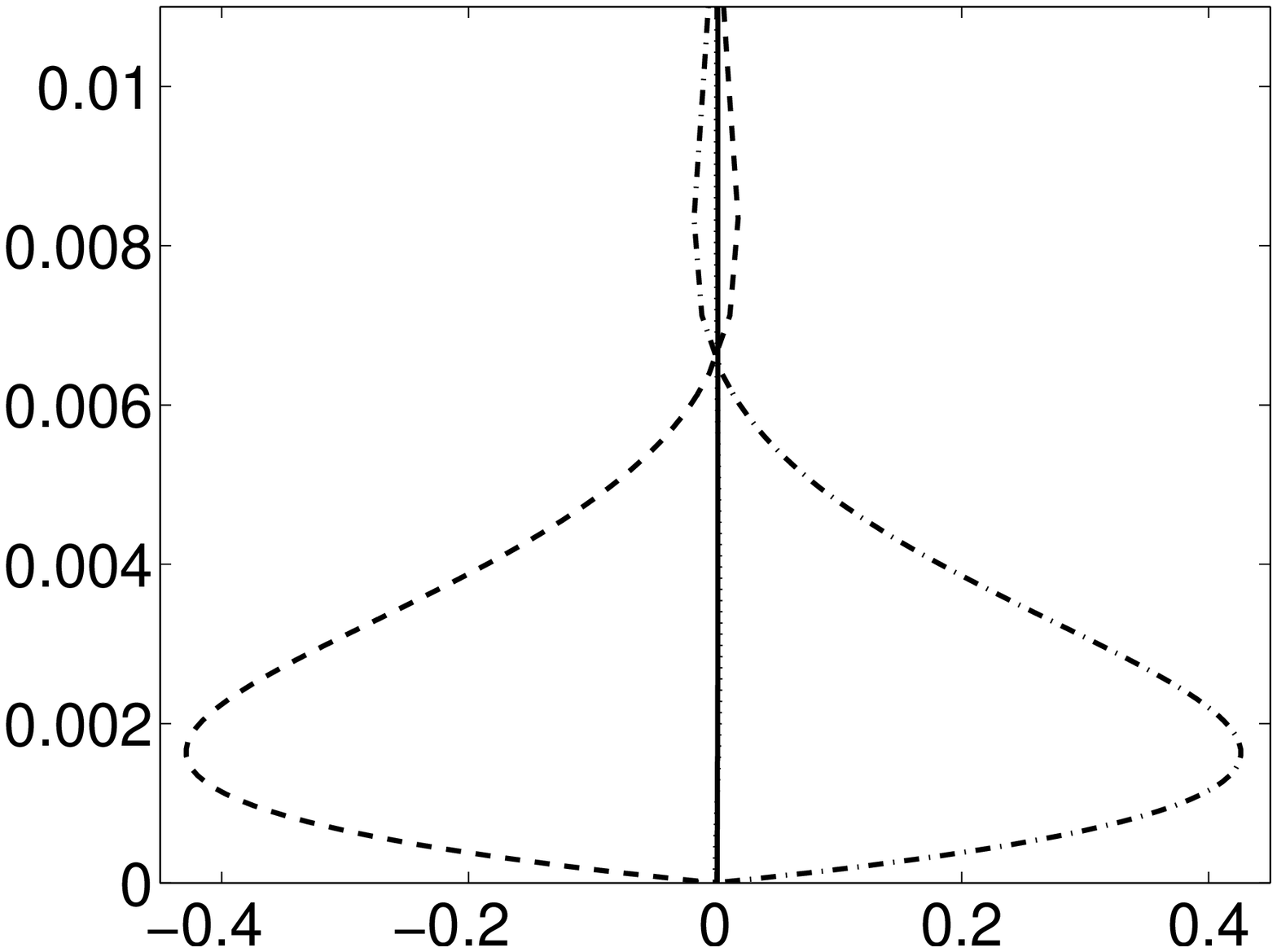 ,width=2in}}
\end{picture}
\end{minipage}
\\ 
\begin{minipage}[t]{3.0in}
\begin{picture}(3.0,2.5)
\centerline{\psfig{figure=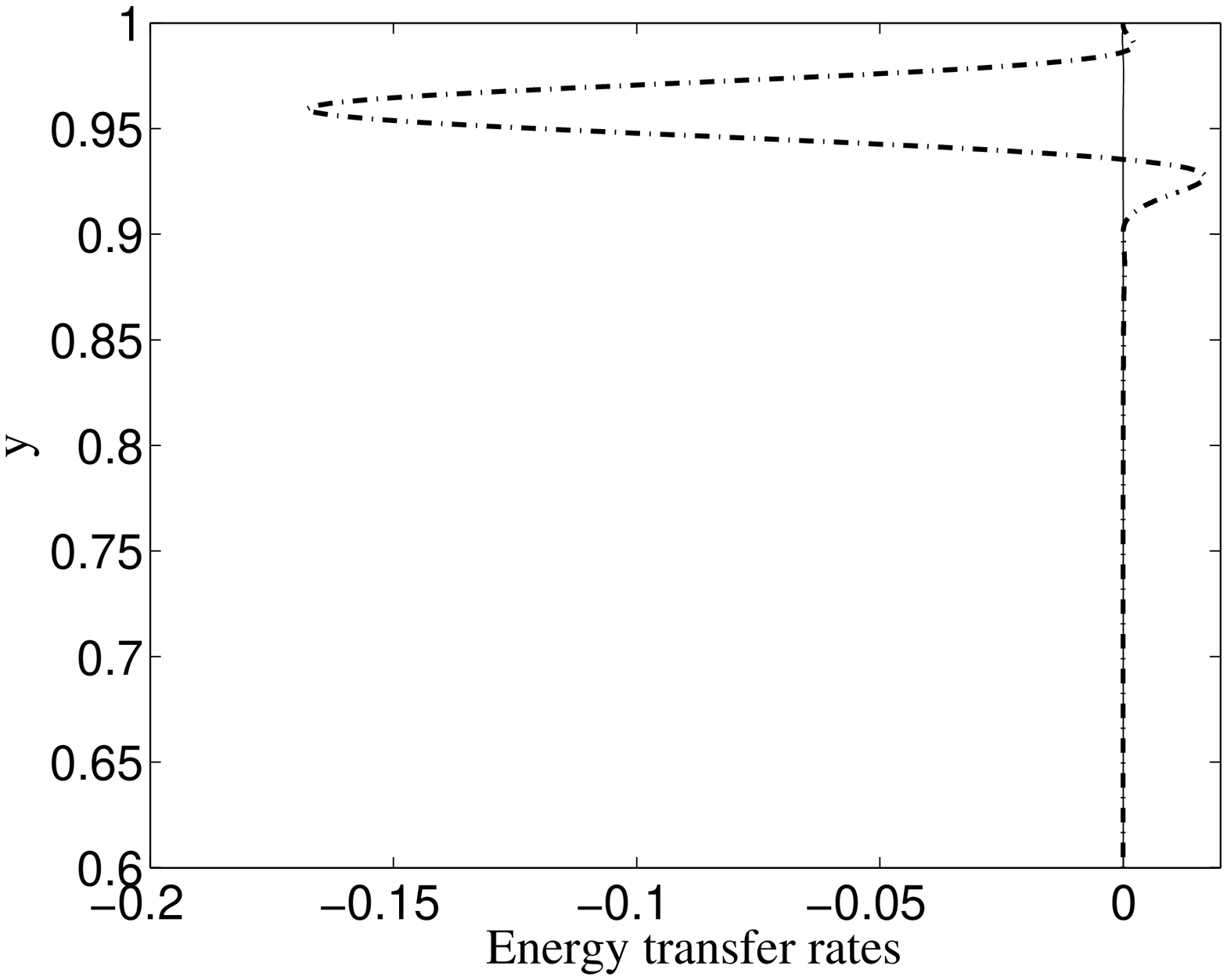 ,width=3.0in}}
\put(-3.0,2.1){(b)}
\put(-2.4,0.3){\psfig{figure=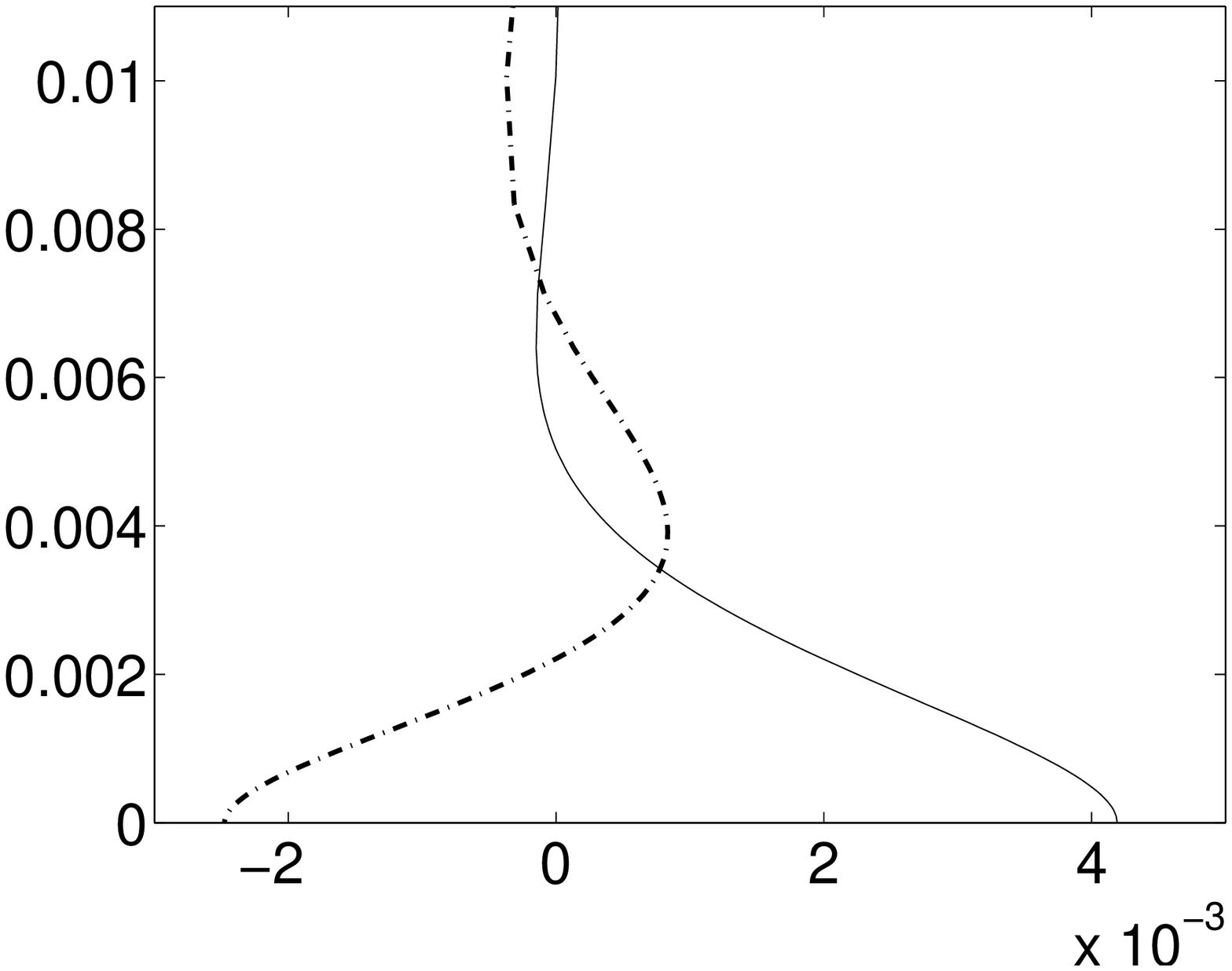 ,width=2in}}
\end{picture}
\end{minipage}
\end{tabular}
\end{center}
\caption{
Energy transfer rates, ${\cal E}_i$, versus $y$ for $\alpha = 2.75$, $\beta = 0$ and 
${\it Re} = 4 \times 10^5$ at $M=3$. (a) Solid line, rate of total energy; 
dashed line, viscous dissipation; dash-dotted line, transfered by pressure terms; 
dotted line, rate of transfer from mean-flow. 
(b) solid line, rate of heat produced by friction, dash-dotted line, 
thermal diffusion rate. Insets in each panel show them at close to the lower wall.
}
\label{fig:fig11}
\end{figure}

\begin{figure}[h!]
\begin{center}
\begin{tabular}{c}
\begin{minipage}[t]{3.0in}
\begin{picture}(3.0,2.5)
\centerline{\psfig{figure=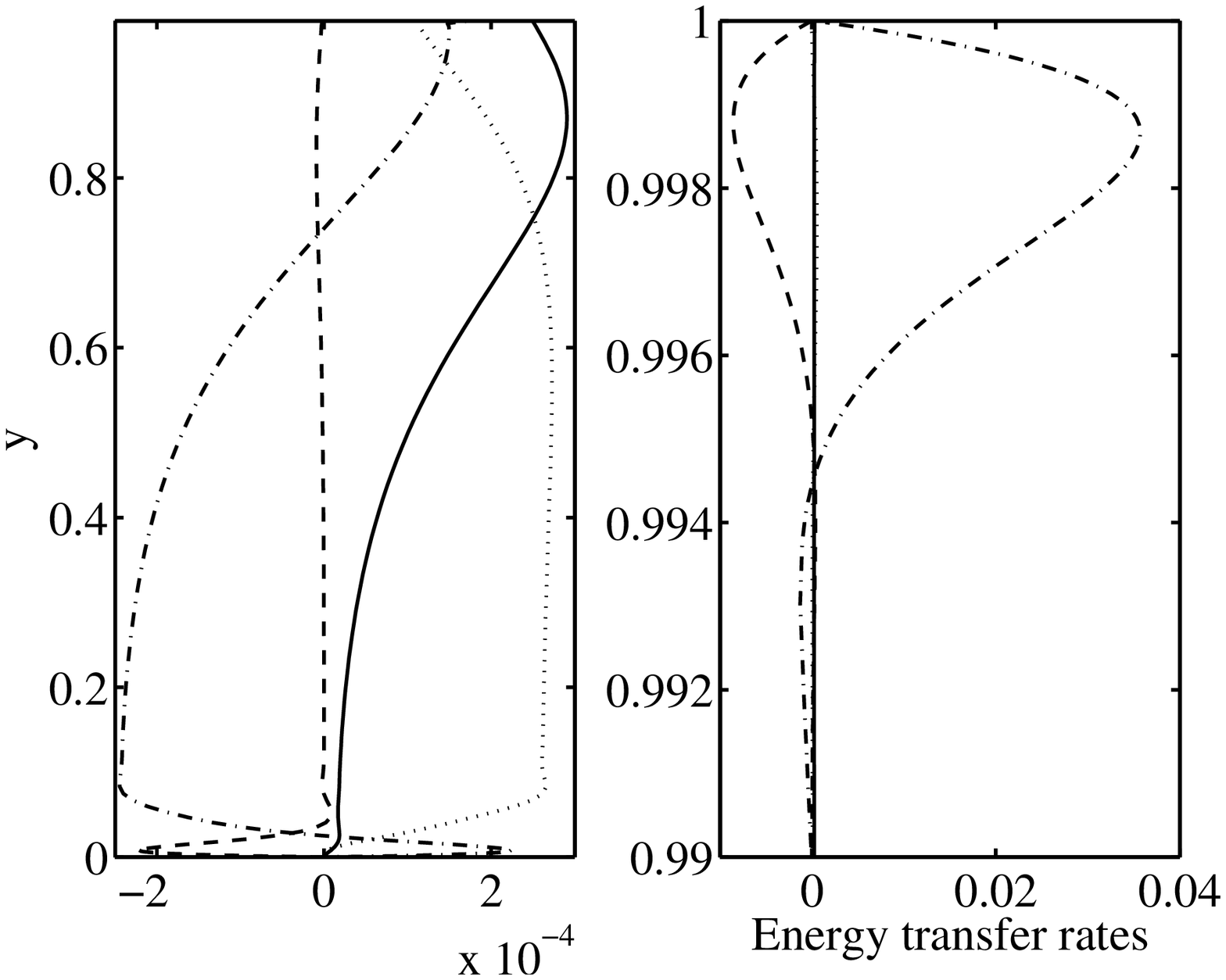,width=3.0in}}
\put(-3.0,2.1){(a)}
\end{picture}
\end{minipage}
\\ 
\begin{minipage}[t]{3.0in}
\begin{picture}(3.0,2.5)
\centerline{\psfig{figure=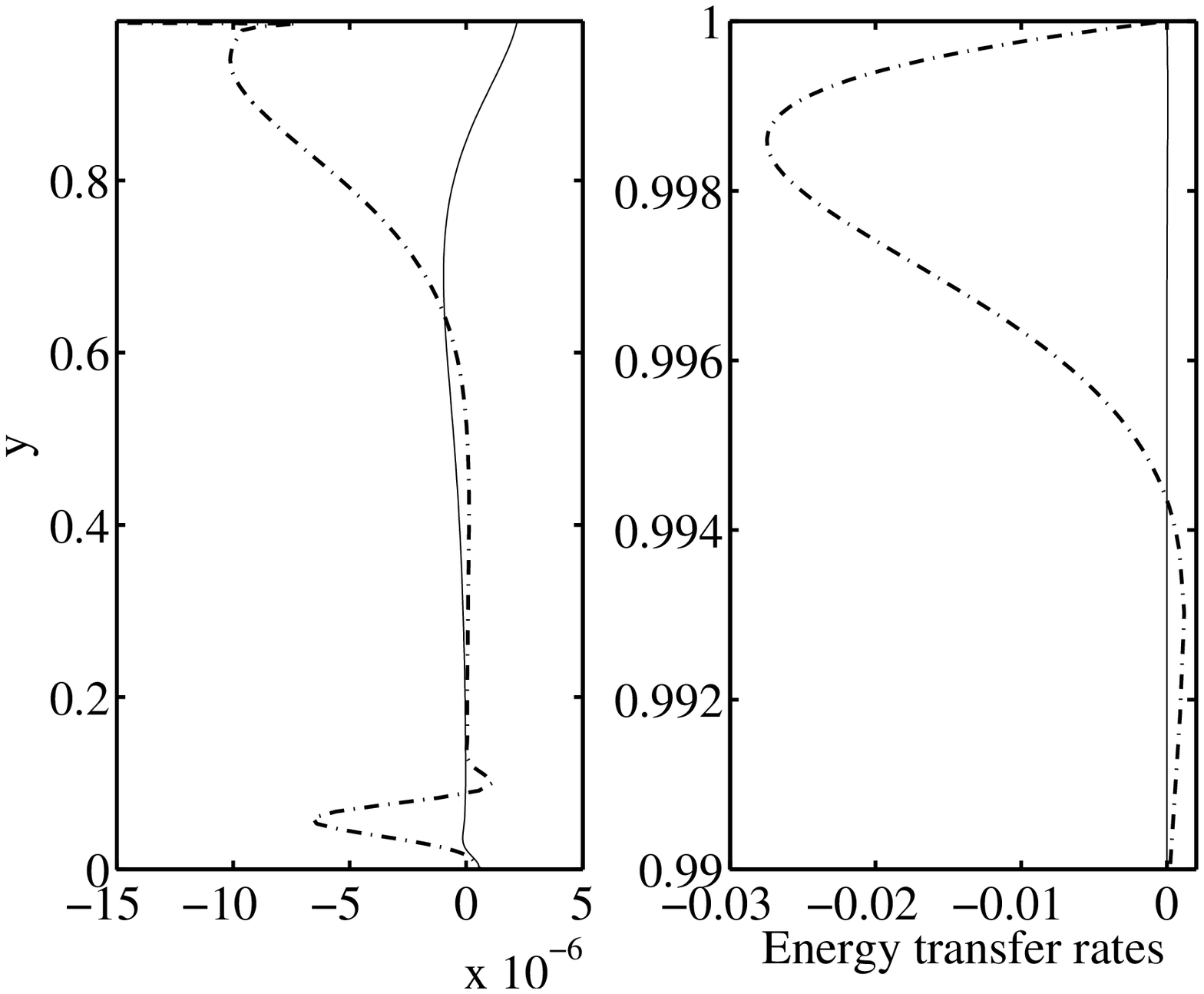,width=3.0in}}
\put(-3.0,2.1){(b)}
\end{picture}
\end{minipage}
\end{tabular}
\end{center}
\caption{
Same as Fig.~11 but for $\alpha = 2.3$. 
The subplots on the right show energy transfer rates near the upper wall.
}
\label{fig:fig12}
\end{figure}

The distinction between mode I and mode II instability becomes
clearer when we look at the distribution of different perturbation energies across $y$.
Figure~11 shows this distribution for $\alpha = 2.75$ 
[which belongs to mode I in Fig.~8(a)], 
and Fig.~12 shows the same for $\alpha = 2.3$  (which belongs to mode II).
These figures  show that the energy-transfer rate from mean-flow 
occurs close to the moving and isothermal top wall for mode I, whereas it occurs 
in the bulk of the flow domain for mode II. 
As one expects the viscous-dissipation is culminated near the walls. 
This is more at the lower wall for mode I, and at the upper wall for mode II. 
Both modes exhibit a larger loss-by-conduction  near the top wall.
This is because the temperature gradient is set to zero at the lower wall via
the adiabatic boundary condition. The heat generated due to viscous dissipation 
is more near the lower wall for mode I and higher near the upper wall for mode II. 
This is in accordance with the momentum loss due to viscous dissipation for both modes. 
Finally, apart from the constituent energy-transfer rates found in appendix, 
Figs.~11 and 12 also show the local energy-transfer rates by pressure. 
The expression for this quantity, say, $\dot{\cal E}_5(t,y)$ is 
\begin{equation}
  \dot{\cal E}_5(t,y) = -\frac{\exp[2\Im(\omega_{ld})t]}{\gamma M^2} 
       (p'^{\dagger} Dv'+v'^{\dagger} Dp') + c.c.
\label{Edot4p}
\end{equation}
Though this quantity does not contribute to the overall total energy-transfer rate
(since this quantity vanishes upon integration across the channel width), 
it plays a role in the distribution of the same across the channel width.

\section{{\label{TranGrowth}}Transient Energy  Growth}

Let us write the linear stability equations in an evolution form:
\begin{equation}
  \frac{\partial {\bf \tilde{q}}}{\partial t} = -{\rm i} {\cal L}{\bf \tilde{q}},
\label{qstateeqn}
\end{equation}
where ${\bf \tilde{q}}(y,t;\alpha,\beta)$ is the inverse Fourier transform of 
$\hat{\bf q}(x,y,z,t)$; the elements of the linear operator, ${\cal L}$,
are omitted for sake of brevity.
In contrast to the {\it modal} linear  stability analysis
that deals with the long-time dynamics of any system via the normal-mode approach,
the key idea of the {\it non-modal}  analysis is 
to probe the short-time dynamics of the system
in terms of perturbation energy in the parameter space 
where the flow is {\it stable} (such as in Fig.~2) according to the 
linear stability analysis,
and investigate the potential of such stable flows to {\it amplify}
the initial perturbation energy.

Let $G(t, \alpha,\beta; Re, M)$ be the maximum possible energy amplification 
at any time $t$, i.e.,
\begin{eqnarray}
   G(t, \alpha,\beta; Re, M) 
    \equiv G(t) &=& 
    \max_{\bf \tilde{q}(0)} \frac{{\cal E}(\alpha,\beta,t)}{{\cal E}(\alpha,\beta,0)} ,
\label{GeqnL2norm}
\end{eqnarray}
where $G(t)$ is optimized over all initial conditions
which is computed using the singular value decomposition.
For an efficient computation of $G(t)$, only a selected portion of the spectra (see Fig.~1)
is chosen~\cite{MAD06}, corresponding to 
the modes whose phase speeds are within the range $-1 < \omega_r/\alpha < 2$
(i.e., comparable to the extremes of the mean flow velocity which varies between $0$ and $1$), 
and the decay rate is less than $0.5$ (i.e., $\omega_i > -0.5$). 
With this choice of modes, the number of selected modes $K$
($<<5N$, where $(N+1)$ is the number of collocation points)
can be reduced by a factor of $5$ or more.  
The related details on numerical scheme are documented in our
earlier paper~\cite{MAD06}.

\begin{figure}[h!]
\includegraphics[width=7.0cm]{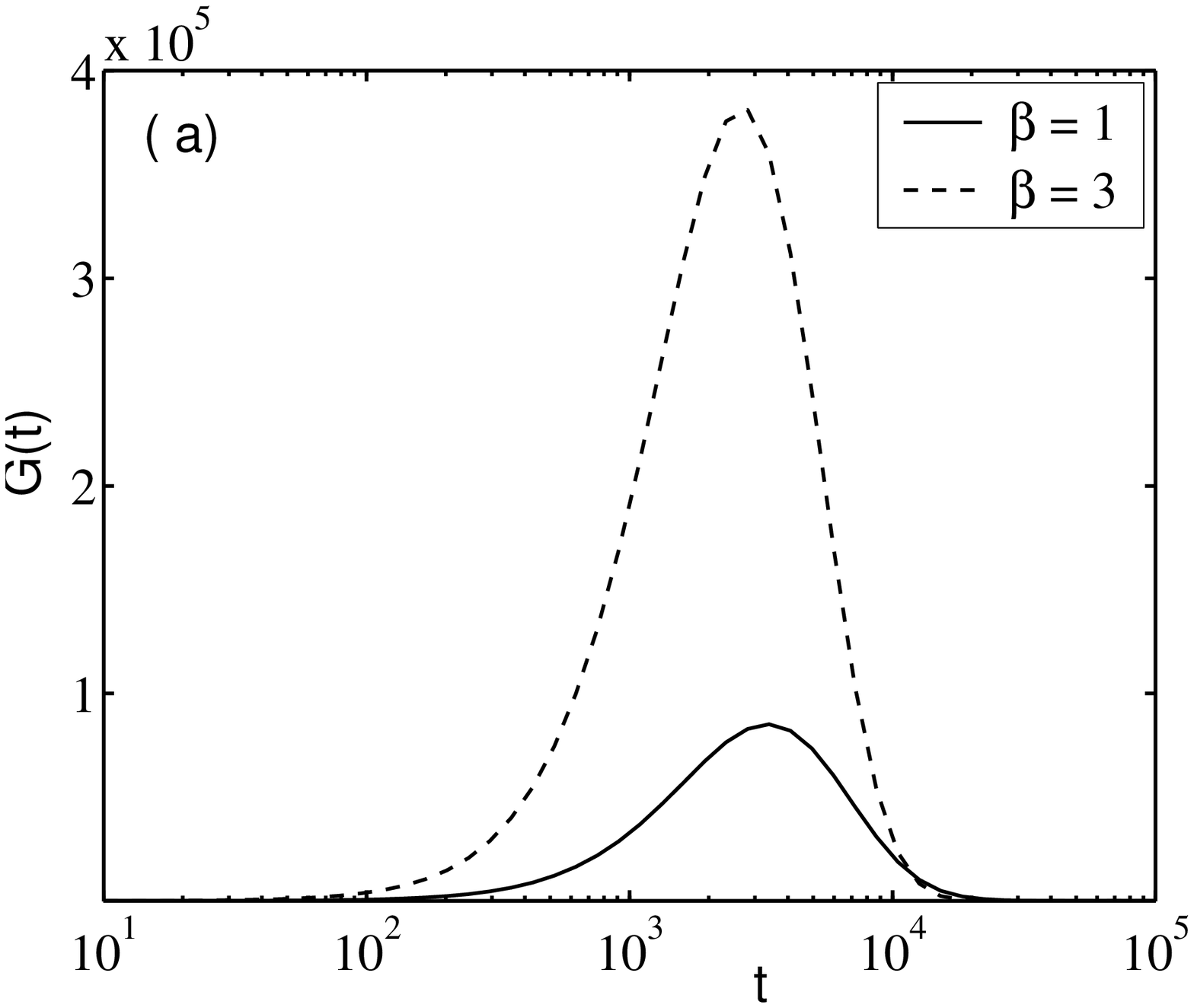}\\
\includegraphics[width=7.0cm]{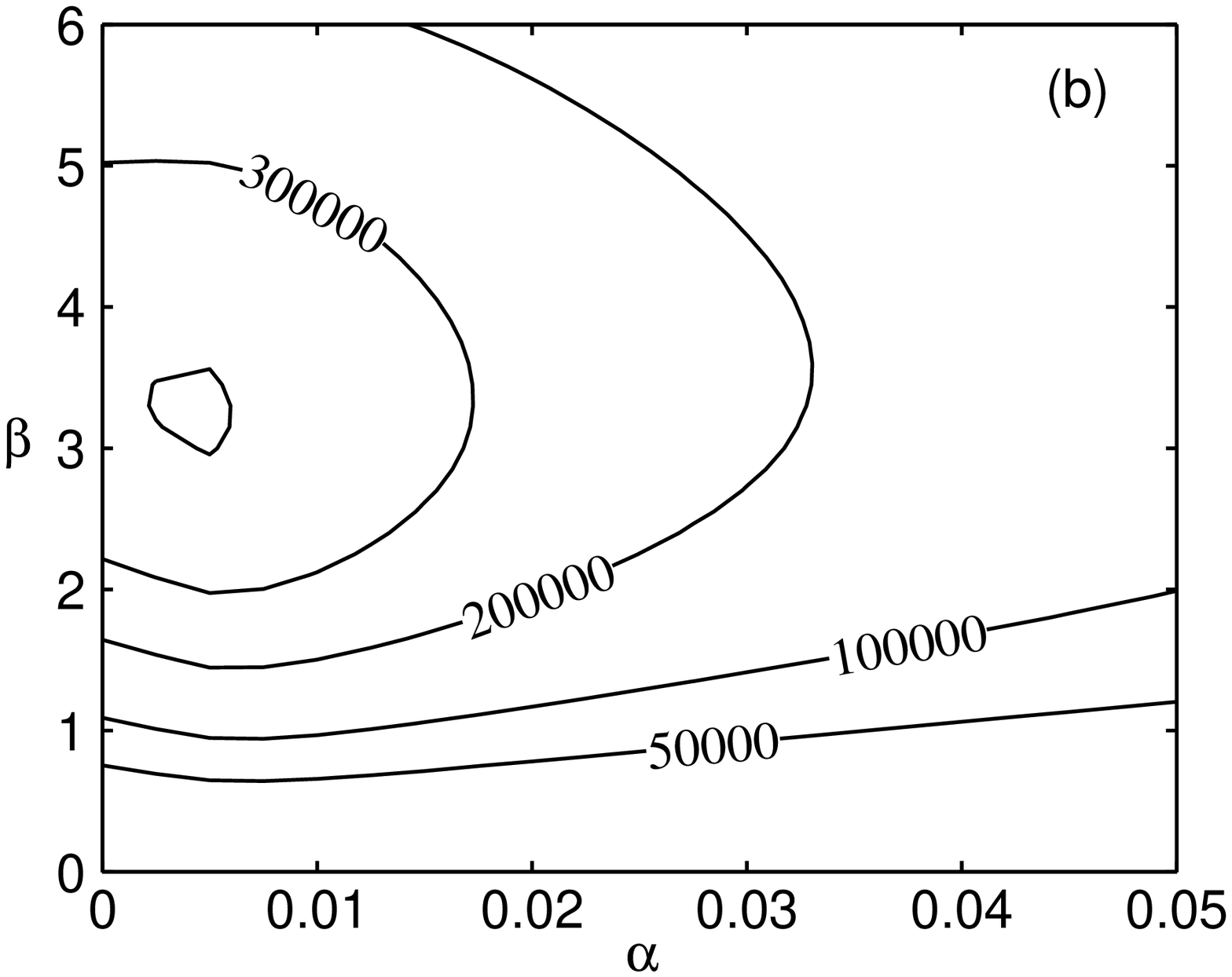}\\
\includegraphics[width=7.0cm]{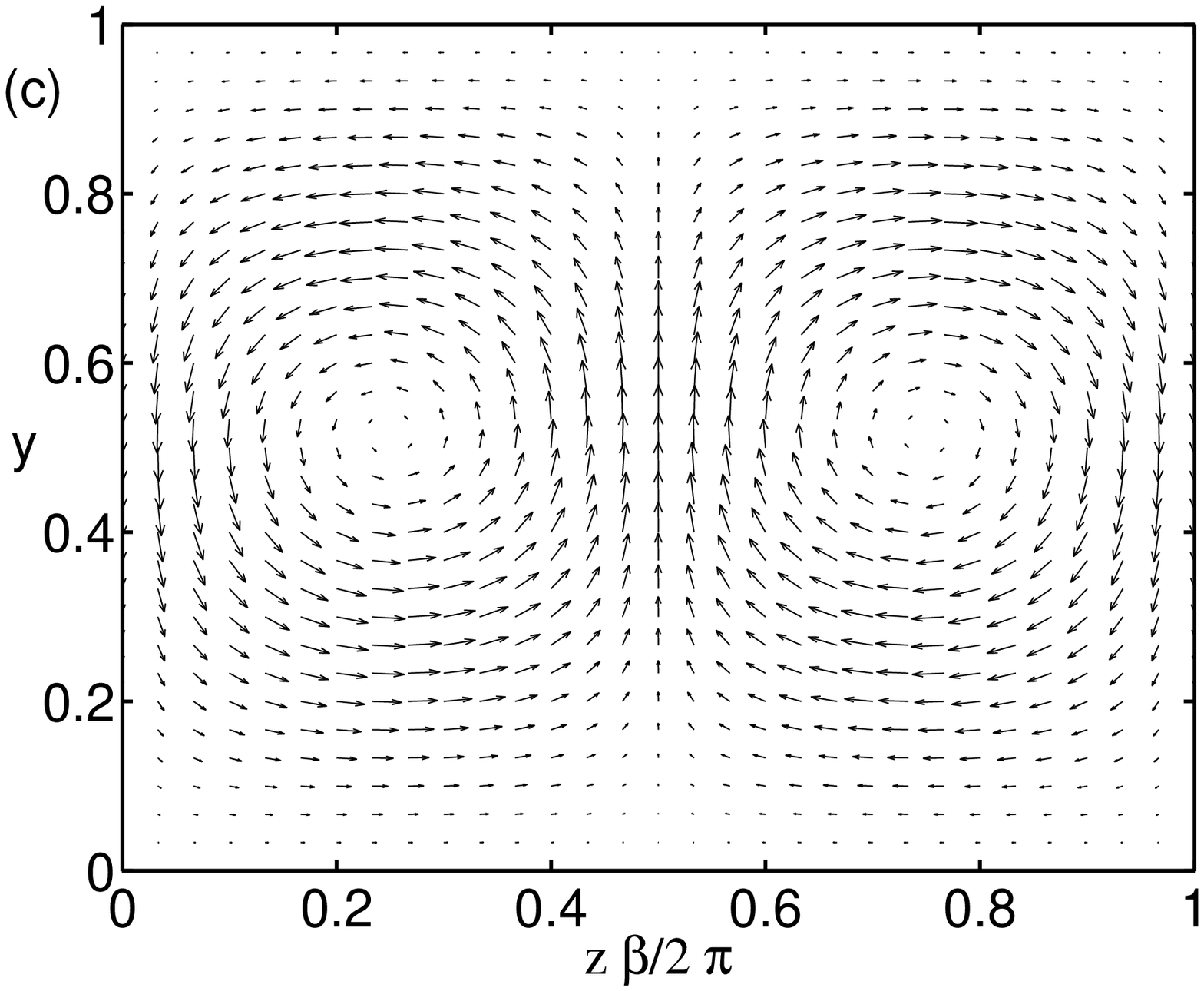}
\caption{
(a) Variation of the energy amplification factor, $G(t)$, with time
for parameter values of ${\it Re} = 10^5$, $M = 2$ and $\alpha = 0$.
(b) Contours of $G_{\max}$ in the ($\alpha, \beta$)-plane
for ${\it Re} = 10^5$ and $M = 2$.
(c) Pattern of optimal perturbation velocities at $t=0$ 
in the ($z, y$)-plane for $\alpha = 0$, $\beta = 3$, $M = 2$ and ${\it Re} = 10^5$.
}
\label{fig:fig13}
\end{figure}

\subsection{Results on Energy  Growth and Optimal Perturbations}

The variation of  $G(t, \alpha,\beta)$ with time for uniform shear flow
is shown in  Fig.~13(a) for different span-wise wavenumber $\beta$, with
${\it Re} =  10^5$, $M = 2$ and $\alpha = 0$;
the solid and dash lines correspond to $\beta=1$ and $3$, respectively.
It is observed that the initial energy density can be amplified by a factor of $10^5$ or more
over a time-scale of order $t=O(10^3)$ for both $\beta$;
in the long-time limit ($t\to \infty$), $G(t)$ decays to zero since the flow is stable.
Figure~13(b) shows the contours of the maximum amplification of energy over all time 
[that occurs at $t=t_{\max}$ such as in Fig.~13(a)] in the ($\alpha, \beta$)-plane,
\begin{equation}
   G_{\max}(\alpha,\beta; Re, M) = \max_{t\ge 0} G(t,\alpha,\beta; Re, M)
\end{equation}
for $Re=10^5$ and $M=2$.
It is seen that larger energy amplification occurs for smaller values of streamwise wavenumber.
For the dash line in Fig.~13(a),
the {\it optimal} velocity patterns in the $(y, z)$-plane at $t=0$ is shown in Fig.~13(c).
[The velocity pattern at $t=t_{\max}$ looks similar to that in  Fig.~13(c).]
This represents a pure streamwise vortex 
which is typical of all shear flows~\cite{BF92,TTRD93,MAD06}.
The structural features of optimal patterns in compressible uniform shear flow
look similar to those in incompressible shear flows.

\begin{figure}[h!]
\includegraphics[width=7.5cm]{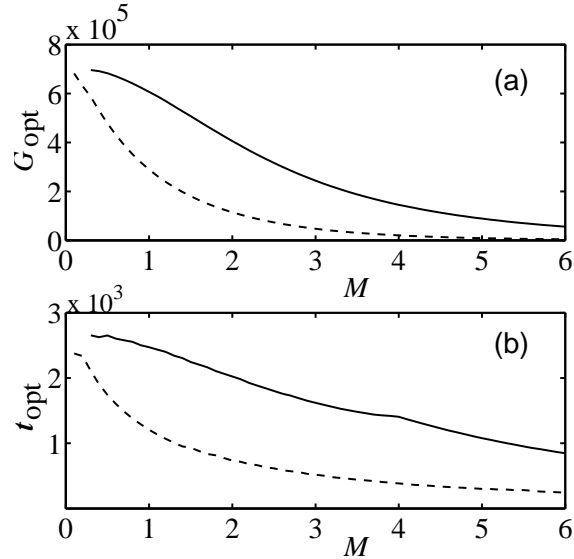}
\caption{
Variations of (a) the optimal energy growth, $G_{\rm opt}$, 
(b) optimal time, $t_{\rm opt}$, with Mach number $M$ for ${\it Re} = 10^5$.
The solid line correspond to the uniform shear mean flow,
and the dashed line to non-uniform shear. 
} 
\label{fig:fig14}
\end{figure}

The global maximum of $G_{\max}$ over all combinations of wavenumber $(\alpha,\beta)$,
\begin{equation}
   G_{\rm opt}(Re, M) = \sup_{\alpha,\beta} G_{\max}(\alpha,\beta; Re, M),
\end{equation}
is called as the {\it optimal energy growth} $G_{\rm opt}$
that occurs at $(t_{\rm opt},\alpha_{\rm opt},\beta_{\rm opt})$. 
The variations of  $G_{\rm opt}$ and the corresponding optimal time $t_{\rm opt}$, 
with Mach number $M$  are shown in Fig.~14(a-b).
The solid and dashed lines in each panel correspond to the uniform
and non-uniform shear flow, respectively; $Re= 10^5$ for these plots. 
Both $G_{\rm opt}$ and $t_{\rm opt}$ decrease monotonically with increasing $M$.
The magnitude of $G_{\rm opt}$ is much
larger for the uniform shear flow; the optimal time $t_{\rm opt}$ is also larger
by a factor of two or more, implying that the energy growth can be sustained
over a longer duration in uniform shear flow. These overall observations on
transient energy growth hold at other sub-critical values of $M$ and $Re$.
Therefore, the uniform shear flow is more susceptible
to sub-critical transitions than its non-uniform counterpart. 
As in the case of modal instability in Section III.C, we can conclude that
{\it the viscosity stratification along with non-uniform shear
would also lead to a ``delayed'' subcritical transition in compressible Couette flow
in terms of nonmodal instability}.

\subsection{{\label{Scaling}}Scalings of $G_{\max}$ and $t_{\max}$}

In a recent paper~\cite{MAD06}, we have shown that the wellknown
scaling law of incompressible shear flows~\cite{Gust91},
{\it $G_{\max}$ varies quadratically with the Reynolds number $Re$,
and $t_{\max}$ varies linearly with $Re$
for streamwise-independent ($\alpha = 0$) modes},
does not hold for the non-uniform shear compressible Couette flow.
To check the validity of this scaling law for the present uniform shear flow,
we have plotted in Fig.~15(a)
the variations of the rescaled energy growth $\sqrt{G(t)}/{\it Re}$
with rescaled time $t/Re$ for four different Reynolds number at $M=2$ and $\beta=1.0$;
the corresponding plot for the non-uniform shear flow is displayed in  Fig.~15(b).
(Plots for different $\beta$ look similar and hence not shown.)
It is clear that the quadratic scaling of $G_{\max}$ with $Re$
holds for the uniform shear case
but does not hold for its non-uniform shear counterpart.

\begin{figure}[h!]
\includegraphics[width=7.5cm]{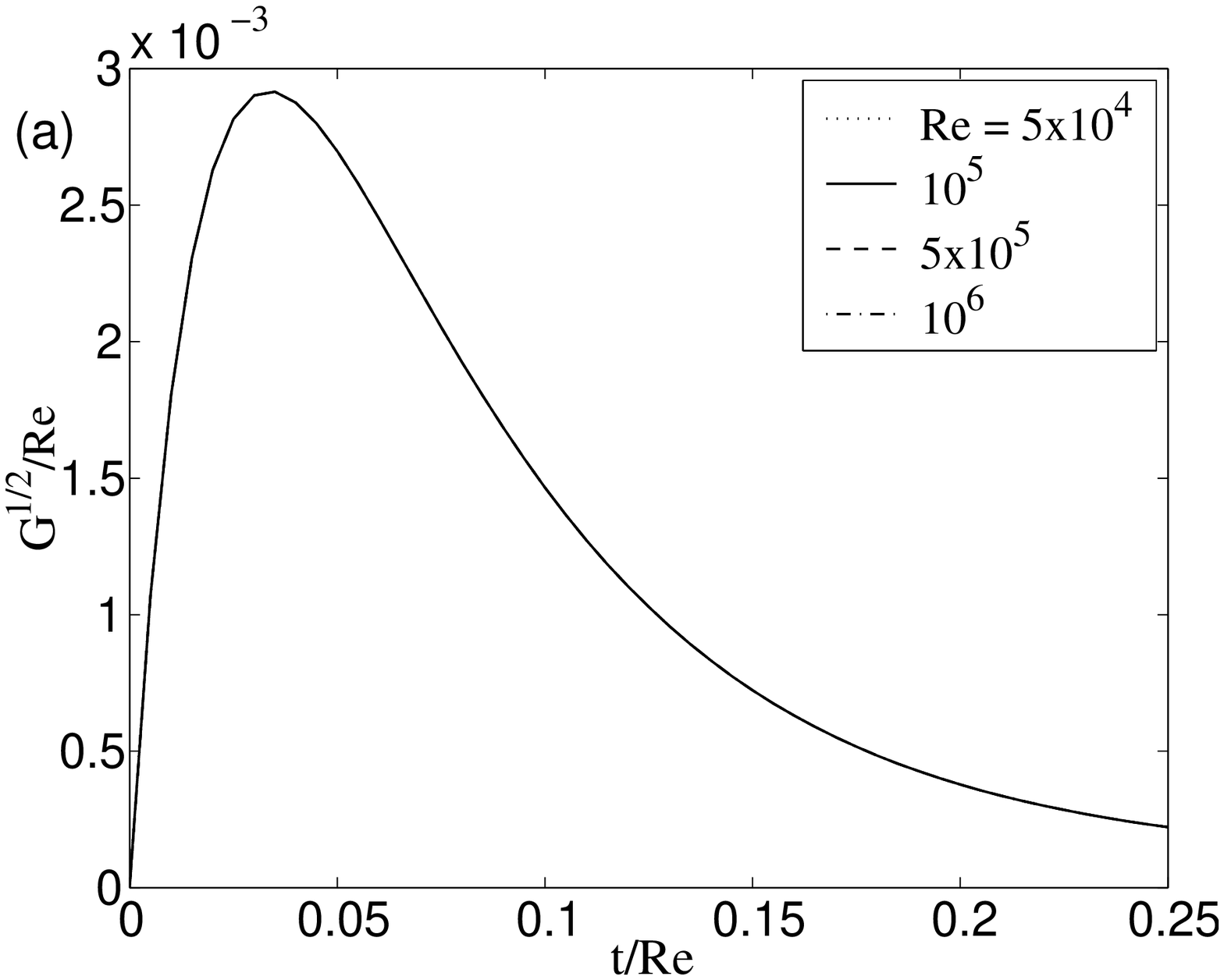}\\
\includegraphics[width=7.5cm]{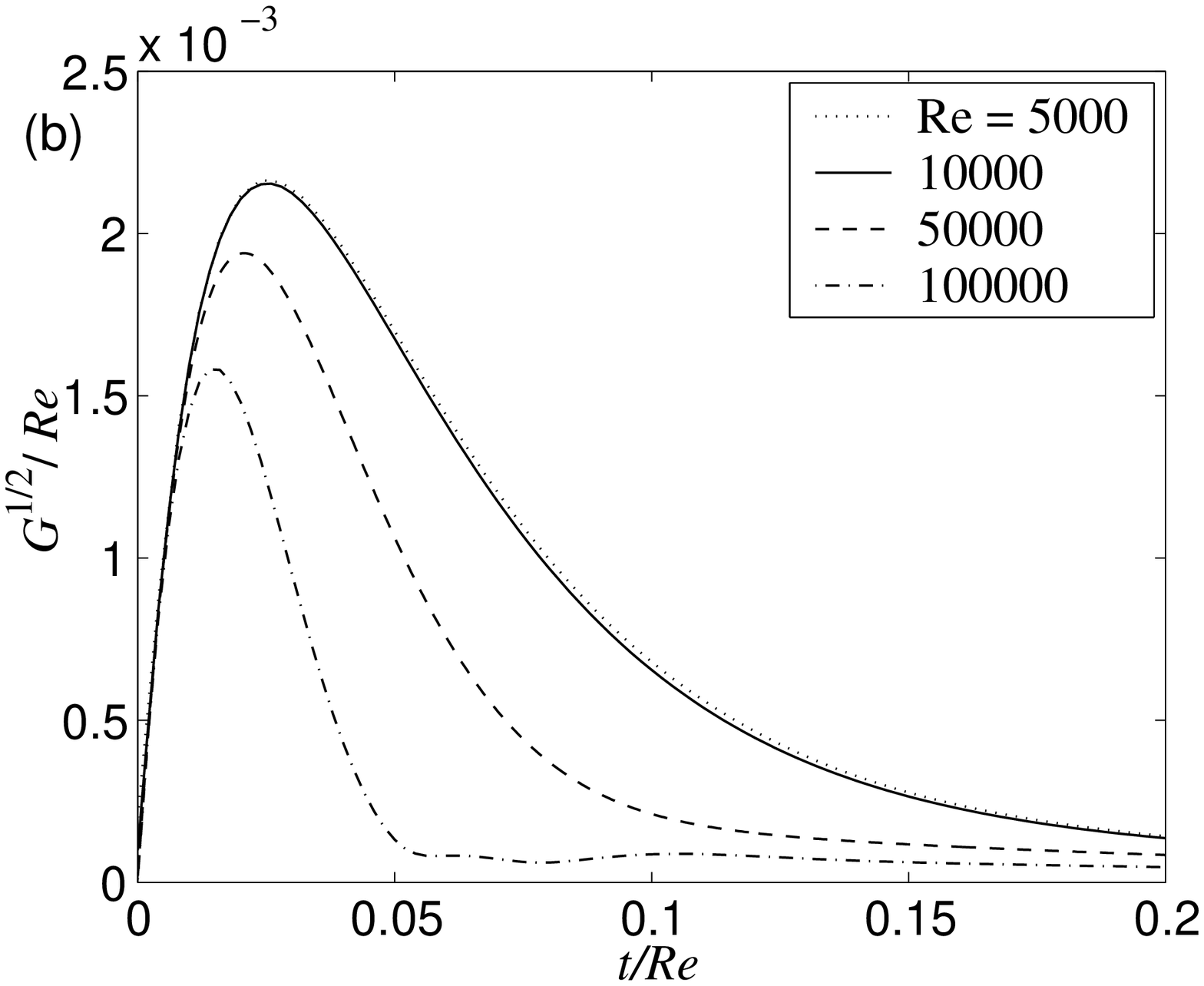}
\caption{
Verification of the  quadratic scaling law for $G(t)$ for
streamwise-independent ($\alpha = 0$) perturbations with $\beta = 1.0$ and  $M = 2$:
(a) Uniform shear; (b) non-uniform shear.
}
\label{fig:fig15}
\end{figure}

For the {\it nonuniform shear flow}, we have argued~\cite{MAD06}
that the following terms, associated with density and temperature fluctuations,
in the $y$- and $z$-momentum equations,
\begin{equation}
\begin{array}{rclrcl}
  {\cal L}_{24} &=& -{\rm i}\left(T_{0y} + T_0 \frac{\rm d}{{\rm d}y}\right)/\rho_0 \gamma M^2, &
  {\cal L}_{34} &=& \beta T_0^2/\gamma M^2, \\
  {\cal L}_{25} &=& -{\rm i}\left(\rho_{0y} + \rho_0\frac{\rm d}{{\rm d}y} \right)/\rho_0 \gamma M^2, &
  {\cal L}_{35} &=& \beta/\gamma M^2,
\end{array}
\label{eqn_L2434}
\end{equation}
are responsible for the violation of the above quadratic scaling-law
since setting them to zero, 
${\cal L}_{24} = {\cal L}_{25} = {\cal L}_{34} = {\cal L}_{35} = 0$,
the rescaled energy-growth curves for different ${\it Re}$ collapses onto a single curve.
Interestingly, for the {\it uniform shear flow} too,  the above terms
${\cal L}_{24}$, ${\cal L}_{25}$, ${\cal L}_{34}$ and ${\cal L}_{35}$
remain  {\it non-zero}, but {\it the quadratic-scaling still holds}.
In this paper, we resolve this apparent contradiction via the following analysis 
of the linear operator in conjunction with the Mack transformation.

Let us rewrite the linear stability equation (22) as
\begin{equation}
   \frac{\partial {\bf \tilde{q}}}{\partial t} =  -{\rm i}{\cal L}^q{\bf \tilde{q}} 
      - {\rm i}{\cal L}^p{\bf \tilde{{\psi}}},
\label{eqn_LSE2}
\end{equation}
where
\begin{eqnarray}
  {\cal L}^{p}_{ij} & = & {\cal L}_{(i+1)(j+3)}, \quad \mbox{for} \quad
    i = 1,2; \ j = 1,2, \label{Lpdef}\\
  {\cal L}^{q} & = & {\cal L}, \; \mbox{with} \; {\cal L}^{q}_{ij} \;=\; 0, 
   \quad \mbox{for} \;  i = 2,3; \ j = 4,5, \label{Lqdef} \\
  {\bf \tilde{{\phi}}} & = & \{\tilde{v}, \tilde{w}\}
    \quad \mbox{and} \quad
      {\bf \tilde{{\psi}}}  \;=\;  \{\tilde{\rho}, \tilde{T}\}. \label{psirhoT}
\end{eqnarray}
Note that the operator ${\cal L}^{p}$ comes from $y$ and $z$-momentum equations,
with elements as in Eq.~(22).
Under the Mack transformation~\cite{Mack84}, $\{\tilde{u}, {\bf \tilde{{\phi}}}, 
{\bf \tilde{{\psi}}}, t\} \rightarrow \{ {\it Re}\,\bar{u}, 
{\bf \bar{{\phi}}}, {\it Re}\,{\bf \bar{{\psi}}}, {\it Re}\,\bar{t}\}$, 
Eq.~(\ref{eqn_LSE2}) transforms into 
\begin{equation}
   \frac{\partial {\bf \bar{q}}}{\partial \bar{t}} =  -{\rm i}\bar{\cal L}{\bf \bar{q}} 
     - {\rm i}{\it Re}^2{\cal L}^p{\bf \bar{{\psi}}},
\label{qbarevolvetbar}
\end{equation}
where $\bar{\cal L}$ is independent of ${\it Re}$ and 
${\bf \bar{q}}=(\bar{u}, \bar{{\phi}}, \bar{{\psi}})^T$. 
In terms of these barred-variables, an evolution equation  
for the total perturbation energy density (15) can be derived as
\begin{equation}
  \frac{\partial \bar{\cal E}}{\partial \bar{t}} =  
   -{\rm i}\int_0^1{\bf \bar{q}}^{\dagger}{\cal M}\bar{\cal L}{\bf \bar{q}} {\rm d}y
    -{\rm i}\ {\it Re}^2 \int_0^1 \rho_0 {\bf \bar{{\phi}}}^{\dagger}
       {\cal L}^p{\bf \bar{{\psi}}} {\rm d}y + c.c. ,
\label{Ebar_evol_tbar}
\end{equation}
where $c.c.$ represents complex conjugate terms.
This equation can be integrated with respect to $\bar{t}$ to yield,
\begin{equation}
   \bar{\cal E}(\bar{t}) = \bar{E}(\bar{t}) + {\it Re}^2 \bar{\cal E}_p(\bar{t}),
\label{Ebartbar_eqn}
\end{equation}
where $\bar{E}(\bar{t})$ is the first term in Eq.~(\ref{Ebar_evol_tbar}) 
integrated with respect to $\bar{t}$, and
the second term, $\bar{\cal E}_p(\bar{t})$, represents the
energy associated with operator ${\cal L}^p$.
If we  divide ${\cal L}^p$ by ${\it Re}^2$
in Eq.~(\ref{qbarevolvetbar}), then Eq.~(\ref{Ebartbar_eqn}) 
becomes independent of ${\it Re}$, and hence we expect the scaling of $G(t)$ to hold.

The above  analysis is verified in Fig.~16
where the energy growth curves for different Reynolds numbers
are seen to collapse on a single curve for the rescaled 
operator ${\cal L}^p\to{\cal L}^p/Re^2$ in Eq.~(\ref{qbarevolvetbar}).

\begin{figure}[h!]
\includegraphics[width=7.5cm]{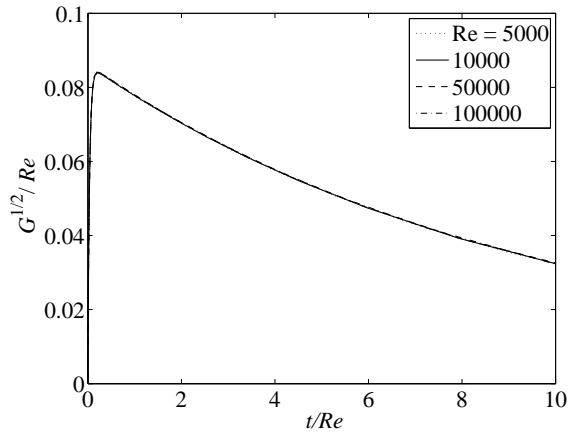}\\
\caption{
Variation of $\sqrt{G}/Re$ with time by rescaling the operator
${\cal L}^p\to{\cal L}^p/Re^2$ in Eq. (31),
with parameters as in Fig. 15(b).
}
\label{fig:fig16}
\end{figure}

It is interesting to note in Fig.~15(b) that the scaling $G(t)\sim Re^2$
holds at low Reynolds numbers ($Re\leq 10^4$) even for the non-uniform shear flow,
and hence the contribution of ${\cal L}^p$ to the perturbation energy must be 
negligible for such low $Re$.
The latter statement can be confirmed if we explicitly compute the contribution of 
energy due to ${\cal L}^p$.
Let us express the total energy density ${\cal E}(t)$ as
\begin{equation}
   {\cal E}(t) = \sum_{l,k} c_l^\dagger c_k \frac{\exp[-{\rm i}(\omega_k-\omega_l^\dagger)t]}
    {\omega_k-\omega_l^\dagger}
   \int_0^1 {\bf q'_l}^{\dagger}{\cal ML}{\bf q'_k} {\rm d}y + c.c.,
\label{Et_repL}
\end{equation}
where $c_k$'s are the expansion coefficients of ${\bf \tilde{q}}$ :
\begin{equation}
   {\bf \tilde{q}}(y,t) = \sum_k c_k \exp[-{\rm i}\omega_k t] {\bf q'_k}(y),
\end{equation}
which can be evaluated by the singular value decomposition of 
the propagator of ${\bf \tilde{q}}$ such that ${\cal E}(t_{\max})=G_{\max}$.
In  Eq.~(\ref{Et_repL}),  the eigenfunction ${\bf q'}$ is  normalized 
(to make the initial total energy ${\cal E}(0) = 1$)
with respect to the weight matrix
$\cal M$, such that $||\tilde{\cal M} {\bf q'_k}|| = 1$, 
where $\tilde{\cal M}$ is given by $\tilde{\cal M}^{\dagger}\tilde{\cal M} = {\cal M}$. 
It is straightforward to verify from Eq.~(\ref{Et_repL})
that the contribution of the terms in Eq.~(\ref{eqn_L2434})
to the total energy is:
\begin{eqnarray}
    {\cal E}_p(t) &=& \sum_{l,k} c_l^\dagger c_k \frac{\exp[-{\rm i}(\omega_k-\omega_l^\dagger)t]}
    {(\omega_k-\omega_l^\dagger)\gamma M^2} \nonumber \\
  & & \hspace*{0.1cm} \times\;
      \int_0^1 [-{\rm i}{v'_l}^{\dagger} Dp'_k + \beta {w_l'}^{\dagger}p'_k]{\rm d}y + c.c.
\end{eqnarray}
Figure~17(a) shows the variation of ${\cal E}_p$  with time at a Reynolds number $Re=10^5$; 
the symbols, circle and triangle, correspond to times at which $G_{\max}$ occurs
for non-uniform and uniform shear flows, respectively.
It is seen that for the case of non-uniform shear 
${\cal E}_p$ at $t = t_{\max}$ is much larger in 
comparison with that for uniform shear. 
At a low Reynolds number $Re=10^4$, however,
${\cal E}_p(t_{\max})$ is negligible for both uniform and non-uniform
shear flows [see Fig.~17(b)], and hence the scaling of $G(t)$ holds
for relatively small  $Re$ [see  Fig.~15(b)] in non-uniform shear flow.

\begin{figure}[h!]
\includegraphics[width=7.5cm]{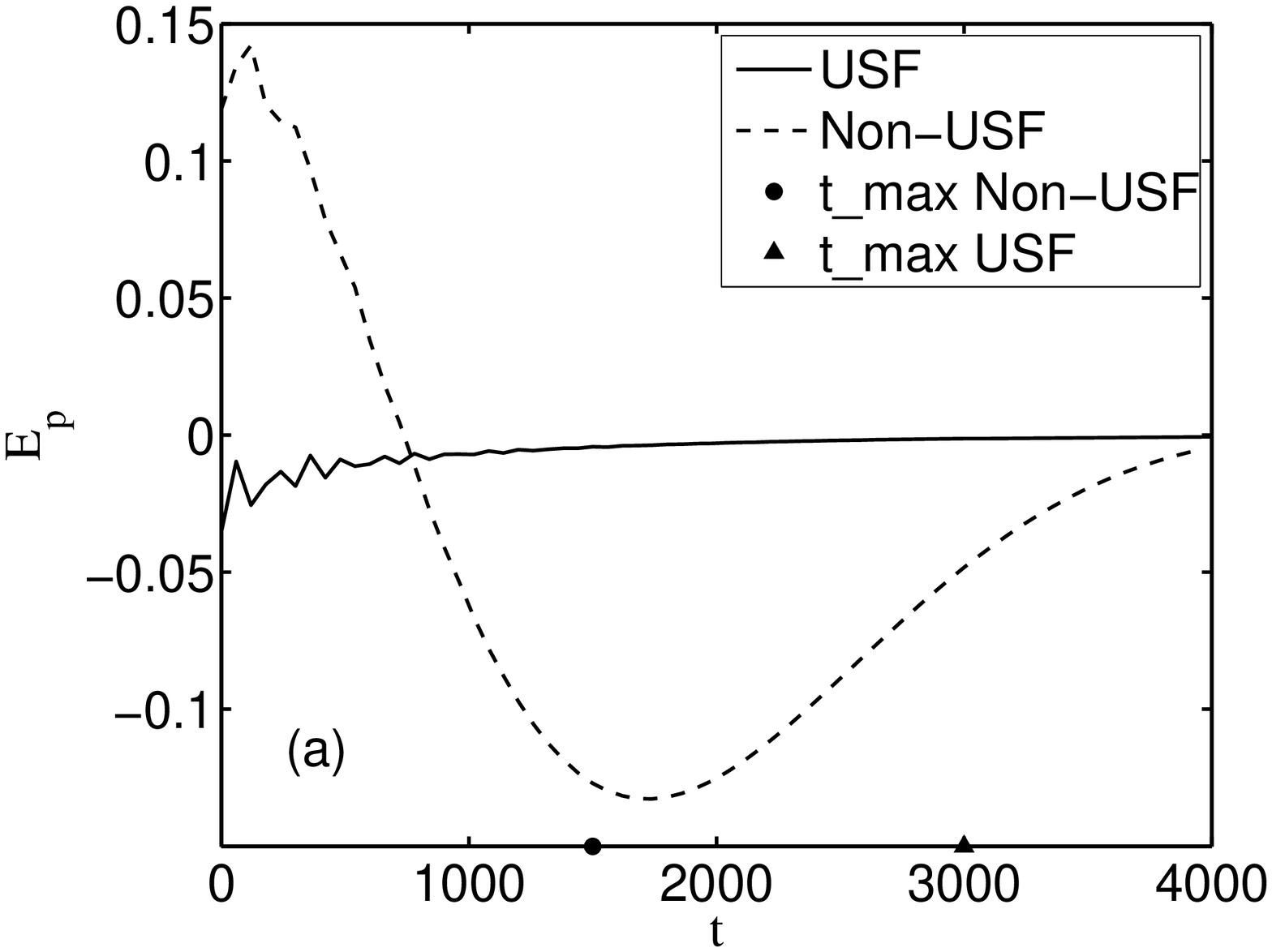}\\
\includegraphics[width=7.5cm]{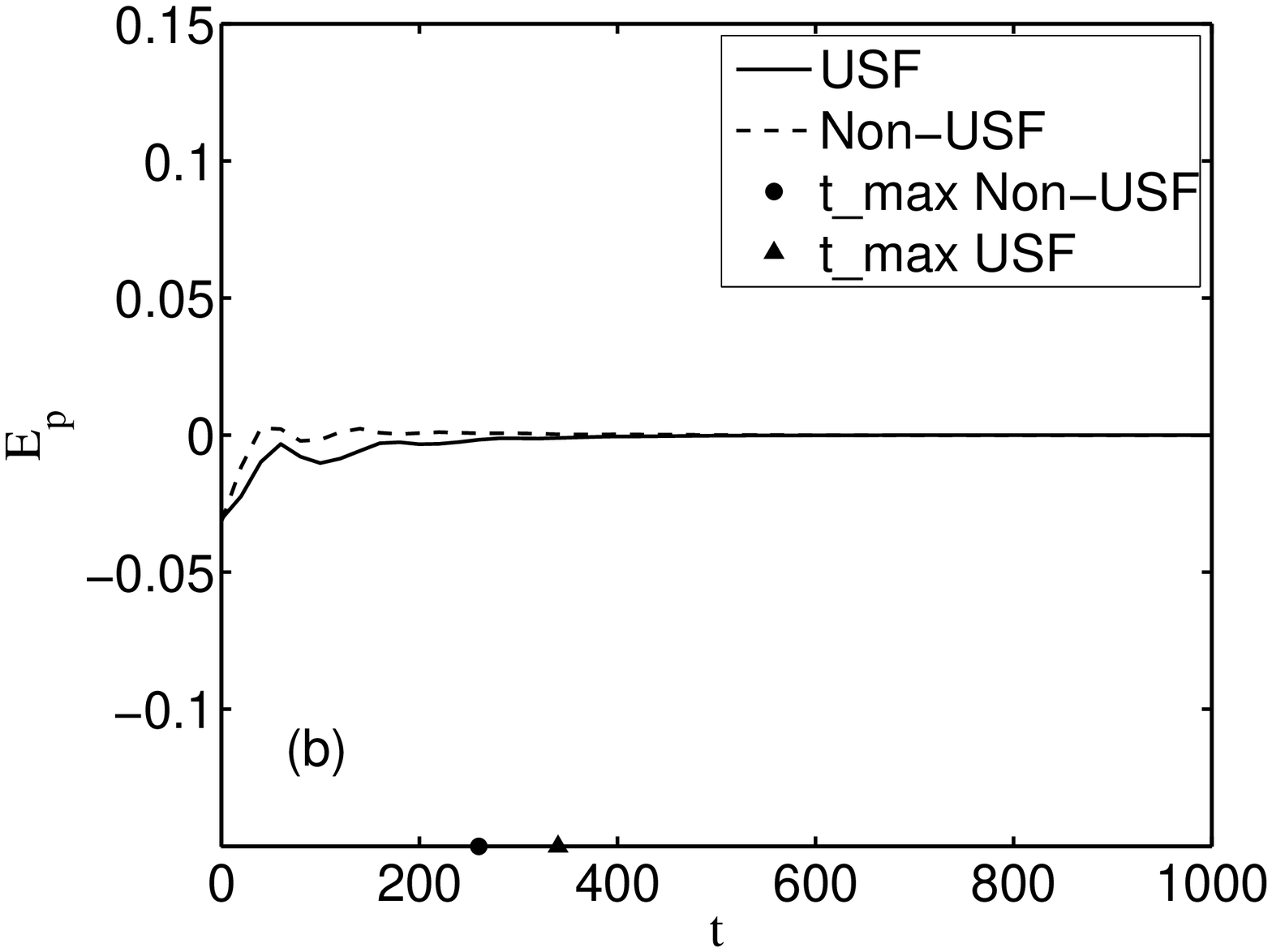}
\caption{
Variation of ${\cal E}_p$ with time for uniform and non-uniform shear flows:
$M = 2$, $\beta = 1$ and $\alpha = 0$.
(a) ${\it Re} =  10^5$ and (b) ${\it Re} =  10^4$. 
}
\label{fig:fig17}
\end{figure}

The above analysis suggests that 
the (streamwise-independent) linear operator $\cal L$ of compressible flows 
can be partitioned into a  Reynolds number dependent operator, ${\cal L}_p$,
and a Reynolds number independent operator $\bar{\cal L}$ 
[Eqs.~(\ref{eqn_LSE2}) and (\ref{qbarevolvetbar})].
The contribution of this $Re$-dependent operator, ${\cal L}_p$, 
to perturbation energy would decide
whether the scaling $G_{\max}\sim Re^2$  would  hold or not for a given  mean flow. 
For the uniform shear flow,  ${\cal L}_p$
has negligible contribution to the energy growth
and hence the quadratic scaling-law holds.

\subsection{Inviscid Algebraic Growth and Optimal Perturbation}

The purely inviscid nature of the algebraic growth suggests one could try to obtain 
the transient growth characteristics directly from inviscid equations. 
As it has been shown numerically in the earlier section 
that the algebraic growth is very pronounced for the modes that are independent of the 
streamwise coordinate (i.e., $\alpha = 0$).  For such an unidirectional flow, 
Ellingsen \& Palm~\cite{EP75} had found an analytical solution for incompressible
flows. An extension of this solution for  density and temperature perturbations was considered 
for the compressible situation~\cite{HH98} which resulted 
in a constraint due to the continuity equation which relates 
spanwise velocity  with normal velocity.
Further, this solution would also result in another constraint which relates 
density and temperature perturbations,
which was not considered before, but is considered here~\cite{MAD08}.  
The Ellingsen-Palm solution for compressible shear flow can be written as
\begin{eqnarray}
  \tilde{u}_{\rm ivs} &=& u'_{\rm ivs} - U_{0y}v'_{\rm ivs}t,
\label{eqn_invisc_u}
\\
  \tilde{v}_{\rm ivs} &=& v'_{\rm ivs},
\label{eqn_invisc_v}
\\
  \tilde{w}_{\rm ivs} &=& \frac{\rm i}{\beta}Dv'_{\rm ivs},
\label{eqn_invisc_w}
\\
  \tilde{\rho}_{\rm ivs} &=& -\rho_0^2 T'_{\rm ivs} -\rho_{0y}v'_{\rm ivs}t
\label{eqn_invisc_rho}
\\
\tilde{T}_{\rm ivs} &=& T'_{\rm ivs} - T_{0y}v'_{\rm ivs}t,
\label{eqn_invisc_T}
\end{eqnarray}
where $u'_{\rm ivs}$, $v'_{\rm ivs}$ and $T'_{\rm ivs}$ are the initial perturbation quantities 
which are to be determined via an optimization procedure; in the following analysis,
the subscript ``${\rm ivs}$'', which refers to ``inviscid solution'', is dropped 
for the sake of simplicity.
The perturbation energy ${\cal E}(t)$ can be written in the basis of the 
quantities $u'$, $v'$ and $T'$, after removing  $w'$ and $\rho'$ 
using the above mentioned constraints, as
\begin{eqnarray}
   {\cal E}(t) &=& \int_0^1\left( \rho_0|\tilde{u}|^2 + \frac{\tilde{v}^\dagger}{\beta^2}[ 
    \rho_0(\beta^2  - D^2) - \rho_{0y}D ]\tilde{v} \right. \nonumber \\ 
    & & \left. + \frac{\rho_0^2}{(\gamma-1)M^2}|\tilde{T}|^2 \right) {\rm d}y.   
\label{energy_invisc_EP_temp}
\end{eqnarray}

Let $\greekvektor{\tilde{\psi}}  =  \{ \tilde{u}, \tilde{v}, \tilde{T} \}^{\rm T}$ 
and $\greekvektor{\psi}'  =  \{ u', v', T'\}^{\rm T}$. Then the above equation can be written as
\begin{equation}
  {\cal E}(t) = \int_0^1  \greekvektor{\psi}'^{\dagger} A^{\dagger} 
           \hat{\mathcal M} A \greekvektor{\psi}' {\rm d}y,
\label{energy_invisc_EP}
\end{equation}
where $\hat{\mathcal M} = {\rm diag}\{ \rho_0, [ \rho_0(\beta^2 - D^2) - \rho_{0y}D ]/\beta^2,
\rho_0^2/(\gamma-1)M^2 \}$, 
and $A$ is a $3 \times 3$ matrix which can be defined by casting 
Eqs.~(\ref{eqn_invisc_u}),~(\ref{eqn_invisc_v}) and~(\ref{eqn_invisc_T}) 
in the form, $\greekvektor{\tilde{\psi}} = A \greekvektor{\psi}'$. 
Now $\hat{G}(t) \equiv \max_{\greekvektor{\psi}'} {\cal E}(t)$ is given by 
\begin{equation}
    \hat{G}(t) = \max(\{\lambda_k\}),
\end{equation}
where ${\lambda_k}$'s are the eigenvalues of the differential equation 
\begin{equation}
   A^{\dagger} \hat{\mathcal M} A \greekvektor{\psi}' = \lambda \hat{\mathcal M} \greekvektor{\psi}'
\label{inviscideigsys}
\end{equation}
with the boundary conditions $v'(0) = v'(1)=0$. 
In contrast to Hanifi \& Henningson's~\cite{HH98} four-variable model, 
this equation (\ref{inviscideigsys}) has only three dependent variables 
and hence called a ``reduced'' model. 
The constraint of vanishing pressure fluctuation is essential to obtain this reduced model; 
the related spatial problem has been solved elsewhere~\cite{MAD08}.

\begin{figure}[h!]
\includegraphics[width=7.5cm]{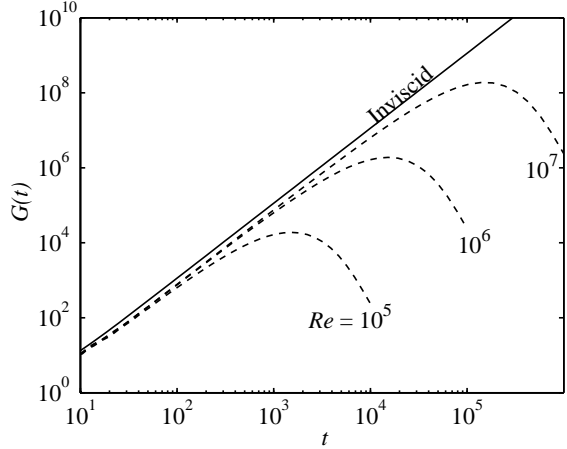}
\caption{
Viscous and inviscid energy growth curves for $\beta = 1$ and $M = 5$. 
Solid line, inviscid solution, $\hat{G}(t)$; dashed line, full viscous solution, $G(t)$.
}
\label{fig:fig18}
\end{figure}

Eq.~(\ref{inviscideigsys}) has been  solved using the 
spectral method. Figure~18 shows the inviscid algebraic 
growth curve $\hat{G}(t)$ at $M = 5$ and $\beta = 1$.
The viscous transient growth curves are also shown for three different Reynolds numbers. 
It is seen that for the entire growth duration the viscous and inviscid growths 
agree quantitatively, demonstrating the inviscid nature of the algebraic growth. 
In terms of energy-transfer-rate, only the following term (see appendix):
\begin{eqnarray}
\dot{\cal E}(t)  &=&  -\int_{0}^{1} \left[\rho_0 U_{0y} 
  \tilde{u}^{\dagger}\tilde{v} + \frac{T_0 \rho_{0y}}{\rho_0 \gamma M^2}
  \tilde{\rho}^{\dagger}\tilde{v} \right. \nonumber \\ 
  && \left. + \frac{\rho_0 T_{0y}}{T_0 \gamma (\gamma-1) M^2}
  \tilde{T}^{\dagger}\tilde{v} \right] {\rm d} y + c.c. 
\label{Edot_invisc}
\end{eqnarray}
survives in the inviscid limit. 
It is clear that the energy transfer from the mean flow occurs via the Reynolds stress 
($\tilde{u}^{\dagger}\tilde{v}$) and the
coupling of the normal perturbation velocity with density 
($\tilde{\rho}^{\dagger}\tilde{v}$) and temperature ($\tilde{T}^{\dagger}\tilde{v}$).
The last two contributions ($\tilde{\rho}^{\dagger}\tilde{v}$
and $\tilde{T}^{\dagger}\tilde{v}$) are unique to compressible flows.
Further, Eqs.~(\ref{eqn_invisc_u}),~(\ref{eqn_invisc_rho}) and~(\ref{eqn_invisc_T}) 
also suggest that this inviscid growth is due to the transfer of energy from mean flow 
to $\tilde{u}$, $\tilde{\rho}$ and $\tilde{T}$ via the fluctuation 
in the normal velocity, $\tilde{v}$. 
The continuity is satisfied by a mere readjustment of $\tilde{w}$ which need not 
grow due to this algebraic growth. 
The growth of $\tilde{u}$ eventually would give rise to {\it streaks}.

\begin{figure}[h!]
\includegraphics[width=7.5cm]{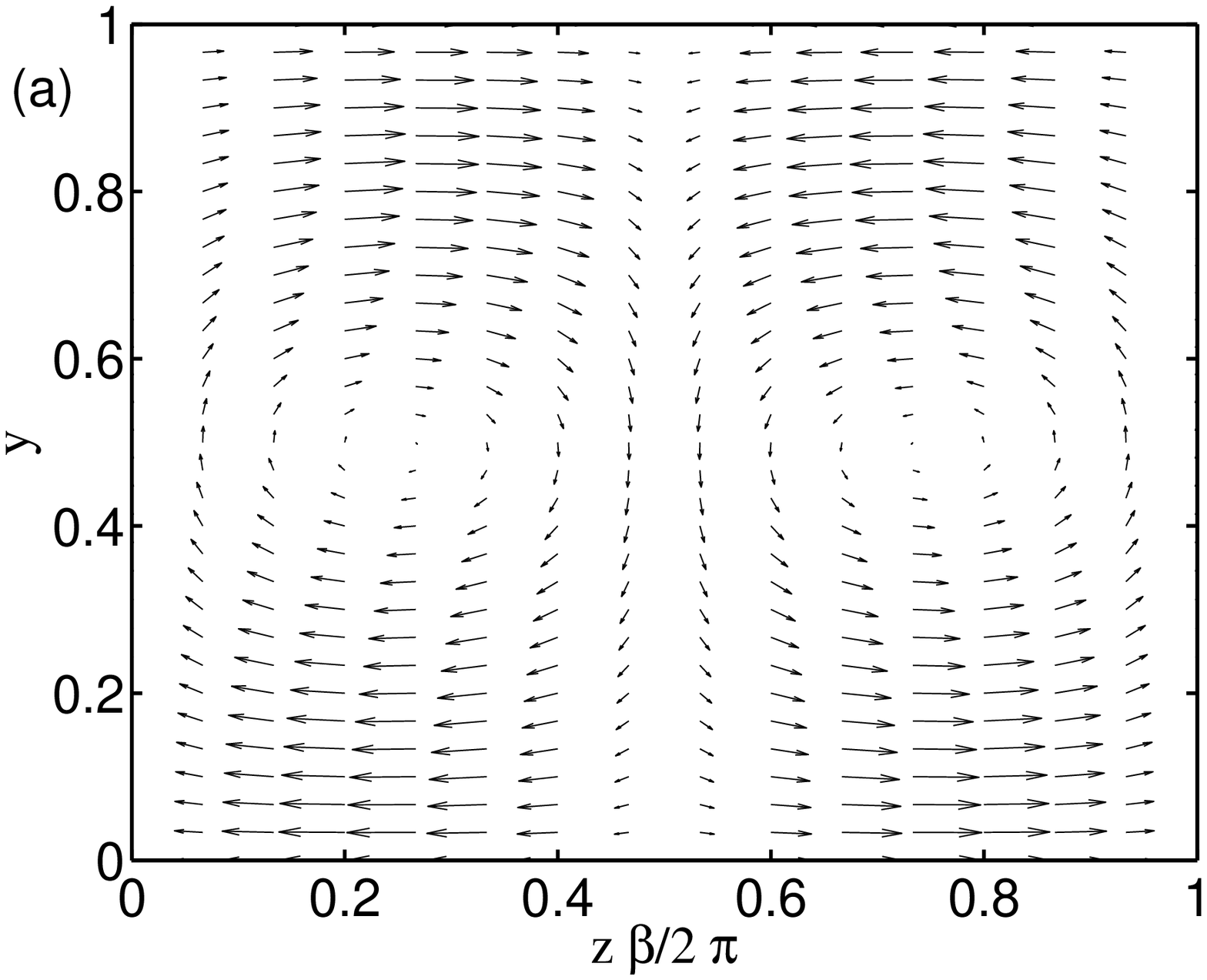}\\
\includegraphics[width=7.5cm]{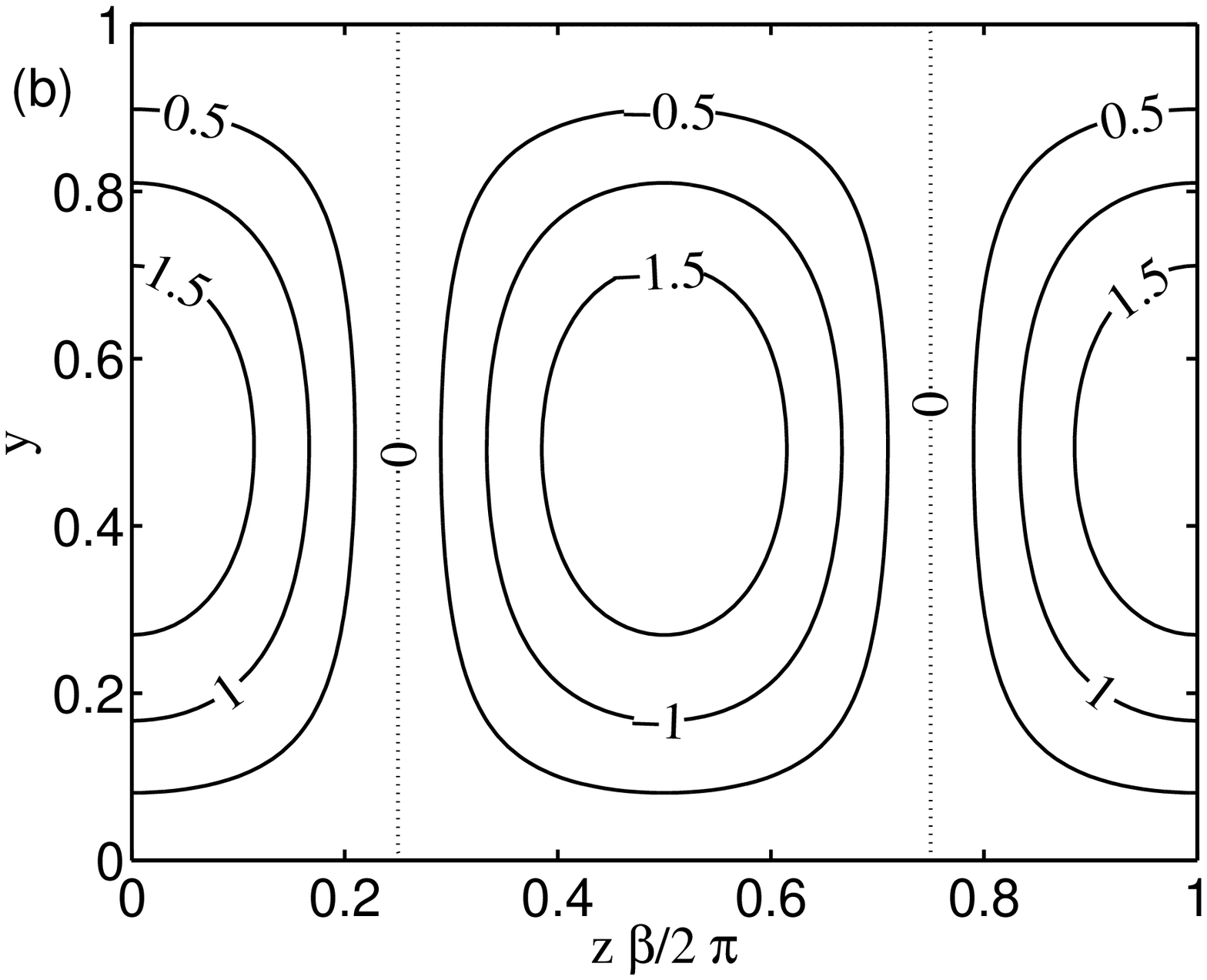}
\caption{
Inviscid optimal patterns of velocity fluctuations given by 
Eq.~(\ref{inviscideigsys}) for $M = 5$ and $\beta = 1$. 
(a) Velocities in $y-z$ plane; (b) contours of $\tilde{u}$.
}
\label{fig:fig19}
\end{figure}

Figure~19 shows the optimal patterns of the perturbation 
velocity-field at $t = 100$, obtained from our reduced inviscid model, Eq.~(45).
Figure~19(a) shows the counter-rotating stream-wise vortices 
in the ($y-z$)-plane, and  Fig.~19(b) shows the contours of 
streamwise velocity fluctuation $\tilde{u}$ 
in the same plane which exhibits the well-known {\it streaks}.  
The structural features of these vortices and streaks are strikingly similar
to those obtained from the solution of full viscous equations. Therefore,
the compressible inviscid Ellingsen-Palm solution, along with the
constraint of null pressure fluctuations, captures all essential features of 
the algebraic growth of the full viscous equations.

\section{{\label{Conclu}}Summary and Conclusion}

The compressible plane Couette flow  is linearly unstable to acoustic disturbances 
for a range of supersonic Mach numbers.
We found that the effects of  viscosity-stratification 
and nonuniform shear rate are to {\it stabilize} the unstable
modes at large stream-wise wavenumber ($\alpha$) and Mach numbers ($M$).
For a given Mach number, the critical Reynolds number ($Re$) 
is found to be significantly smaller (by a factor of 3 or more)
in the uniform shear flow than in its nonuniform shear counterpart;
for a given $Re$, the maximum growth rate (over all $\alpha$, cf. Fig.~2) could be larger
by an order-of-magnitude in the former.
This strong stabilization effect appears to be tied to the
strong viscosity stratification in non-uniform shear flow,
and, therefore, {\it the viscosity stratification would lead to a delayed
transition in compressible Couette flow}. 
Three-dimensional modes could be more unstable than their two-dimensional 
counterparts for some values of $\alpha$, and hence Squire's theorem is, 
in general, not valid for the ``nonisothermal'' compressible Couette flow.
It is shown that the mode II remains the {\it dominant} instability
(i.e., the mode having the maximum growth-rate over all $\alpha$, Eq. 14)
for all Mach numbers in the uniform shear flow.
In contrast, for the nonuniform shear flow, the mode I is
the {\it dominant} instability for low Mach numbers
and the mode II for moderate-to-large Mach numbers.
For both mean flows, the viscosity plays the dual role
of destabilizing (at small $\alpha$) and stabilizing (at moderate-to-large $\alpha$)
the mode II instability, but it destabilizes the mode-I instability.
The higher-order odd (III,...) and even (IV,...) inviscid modes could
also become unstable, but they remain subdominant with respect to
mode I and mode II instabilities.

An analysis based on the perturbation kinetic energies transfered by different terms 
of the governing equation has been carried out to understand the origin of modal instabilities. 
The instability is primarily caused by an excess transfer of energy from mean-flow 
to perturbations for a band of stream-wise wavenumbers. 
It is found that the energy-transfer rate from the mean-flow 
occurs close to the moving and isothermal top-wall for mode I, 
whereas it occurs in the bulk of the flow domain for mode II. 
For 2D modes, the thermal-diffusion process tends 
to stabilize the fluctuations at a higher rate than the viscous dissipation; 
for 3D modes, however, the viscous dissipation
dominates over thermal diffusion at high spanwise wavenumber.

For the transient growth analysis, it is shown that the maximum temporal growth
of perturbation energy, $G_{\max}$, and the corresponding time scale
to attain this maximum, $t_{\max}$, are much larger 
(and can differ by a factor of $5$ or more) for the uniform shear flow
in comparison with the nonuniform shear flow.  
(In other words, the viscosity stratification has a strong stabilizing effect
on transient energy growth.)  Therefore, the uniform shear flow 
is more susceptible to {\it subcritical} transitions than its nonuniform shear counterpart.
For both mean flows,
the optimal energy growth, $G_{\rm opt}$, (i.e., the global maximum of $G_{\max}$ 
in the $(\alpha, \beta)$-plane for given $Re$ and $M$) decreases with increasing $M$; 
pure streamwise vortices ($\alpha_{\rm opt}=0$) are the optimal velocity patterns
at large $M$ but the modulated streamwise vortices ($\alpha_{\rm opt}\neq 0$)
are optimal patterns for low-to-moderate values of $M$.
The physical mechanism of transient energy growth is tied  to
the transfer of energy from the mean flow to perturbations via 
the Reynolds stress and the coupling of density and temperature
perturbations with the normal velocity.

For the streamwise independent perturbations ($\alpha=0$),
we have found that the transient energy growth follows the well-known
scaling law, $G_{\max} \sim {\it Re}^2$ and $t_{\max} \sim  {\it Re}$,
of incompressible shear flow~\cite{Gust91}.
This is in stark contrast to the result on the nonuniform shear flow
for which the above scaling law does not hold~\cite{MAD06}.
An analysis of the linear stability operator, ${\cal L}$, shows that
${\cal L}$ can be partitioned into a $Re$-dependent operator, ${\cal L}_p$,
and a $Re$-independent operator, $\bar{\cal L}$, 
(Eqs.~27 and 31) via the Mack transformation.
The (in)validity of the above scaling laws 
for the (non-)uniform shear flow is shown to be tied   
to the (non-)negligible contribution (to perturbation energy) of ${\cal L}_p$.
Lastly, a `reduced' inviscid model (Eq.~45), based on the inviscid Ellingsen-Palm-type solution, 
has been derived which captures all salient features of 
transient energy growth of full viscous equations.

\begin{appendix}
 
\section{Evolution Equation of Perturbation Energy}

It can be verified  that the perturbation energy
${\cal E}(\alpha,\beta,t)$ satisfies the following time-evolution equation~\cite{MAD06}:
\begin{equation}
  \frac{\partial \cal E}{\partial t} = -{\rm i} \int_{0}^{1}{\bf \tilde{q}^{\dagger} 
    {\mathcal M} {\cal L} \tilde{q}} \ {\rm d}y + c.c.
  = \dot{\cal E}_0 +\dot{\cal E}_1 + \dot{\cal E}_2 + \dot{\cal E}_3 + \dot{\cal E}_4, 
\label{Eevolv} 
\end{equation}
The constituent energy transfer rates, $\dot{\cal E}_0$--$\dot{\cal E}_4$,
have following forms (with $D={\rm d}/{\rm d}y$):
\begin{equation}
 \dot{\cal E}_0  =  -{\rm i}\alpha \int_{0}^{1}U_0 {\bf \tilde{q}^{\dagger} 
    {\mathcal M} \tilde{q}} {\rm d}y + c.c.
  \label{Edot0} 
\end{equation}
\begin{eqnarray}
\dot{\cal E}_1  &=&  -\int_{0}^{1} \left[\rho_0 U_{0y}
  \tilde{u}^{\dagger}\tilde{v} + \frac{T_0 \rho_{0y}}{\rho_0 \gamma M^2}
  \tilde{\rho}^{\dagger}\tilde{v} \right. \nonumber \\
  & &  \left. +\frac{\rho_0 T_{0y}}{T_0 \gamma (\gamma-1) M^2}
  \tilde{T}^{\dagger}\tilde{v} \right] {\rm d} y + c.c.
\label{Edot1}
\end{eqnarray}
\begin{eqnarray}
\dot{\cal E}_2 & = & -\frac{1}{\it Re}\int_{0}^{1} \left[\alpha^2
  (\mu_0 + \lambda_0) \tilde{u}^{\dagger}\tilde{u} + \mu_0(\alpha^2 + \beta ^2)
  \tilde{u}^{\dagger}\tilde{u} 
  \right . \nonumber \\&& \left .
  - \tilde{u}^{\dagger}(\mu_{0y}D + \mu_0D^2)\tilde{u}
  -{\rm i}\alpha\tilde{u}^{\dagger}(\mu_{0y} 
  \right . \nonumber \\&& \left .
   + (\mu_0 + \lambda_0)D)\tilde{v}
  + \alpha \beta (\mu_0 + \lambda_0) \tilde{u}^{\dagger}\tilde{w}
  \right . \nonumber \\&& \left .
  -(U_{0yy}\mu_T + U_{0y} T_{0y}\mu_{TT}) \tilde{u}^{\dagger}\tilde{T}
  -U_{0y}\mu_T \tilde{u}^{\dagger} D\tilde{T}
  \right . \nonumber \\&& \left .
  -{\rm i}\alpha\tilde{v}^{\dagger}(\lambda_{0y} + (\mu_0 + \lambda_0)D)\tilde{u}
  \right . \nonumber \\&& \left .
  + \mu_0(\alpha^2 + \beta^2)\tilde{v}^{\dagger}\tilde{v}
  -\tilde{v}^{\dagger}((\lambda_{0y} + \mu_{0y})D
  \right . \nonumber \\&& \left .
  + (\lambda_0 + \mu_0)D^2 +\mu_{0y} D + \mu_0 D^2)\tilde{v}
  \right . \nonumber \\&& \left .
  - {\rm i}\beta(\lambda_0 + \mu_0) \tilde{v}^{\dagger}D \tilde{w}
  -{\rm i}\alpha U_{0y}\mu_T \tilde{v}^{\dagger} \tilde{T}
  \right . \nonumber \\&& \left .
  -{\rm i}\beta \lambda_{0y} \tilde{v}^{\dagger} \tilde{w}
  -{\rm i}\beta \mu_{0y} \tilde{w}^{\dagger} \tilde{v}
  \right . \nonumber \\&& \left .
  + \alpha \beta (\mu_0 + \lambda_0) \tilde{w}^{\dagger}\tilde{u}
  - {\rm i}\beta(\lambda_0 + \mu_0) \tilde{w}^{\dagger}D \tilde{v}
  \right . \nonumber \\&& \left .
  + (\mu_0(\alpha^2 + \beta^2) + \beta^2(\lambda_0 + \mu_0))\tilde{w}^{\dagger}\tilde{w}
  \right . \nonumber \\&& \left .
  -\mu_0\tilde{w}^{\dagger}D^2\tilde{w} -\mu_{0y}\tilde{w}^{\dagger}D\tilde{w} \right ] {\rm d} y + c.c.
\label{Edot2}
\end{eqnarray}
\begin{eqnarray}
\dot{\cal E}_3 & = & \frac{1}{\sigma{\it Re}(\gamma -1) M^2}
   \int_{0}^{1} \rho_0\tilde{T}^{\dagger}\left[ \mu_T T_{0yy} + T_{0y}^2 \mu_{TT}
     \right. \nonumber \\
    &+& \left. 2T_{0y} \mu_T D -(\alpha^2+\beta^2)\mu_0
     + \mu_0 D^2 \right ]\tilde{T} {\rm d} y + c.c.
\label{Edot3}
\end{eqnarray}
\begin{eqnarray}
\dot{\cal E}_4  &=&  \frac{1}{Re}\int_{0}^{1} \rho_0 \left [
   2\mu_0 U_{0y}\tilde{T}^{\dagger} D \tilde{u} \right. \nonumber \\
   &+& \left.  2{\rm i}\alpha \mu_0 U_{oy}\tilde{T}^{\dagger}\tilde{v}
  + U_{0y}^2 \mu_T \tilde{T}^{\dagger}\tilde{T}
   \right ] {\rm d} y + c.c.
   \label{Edot5}
\end{eqnarray}
Here, $\dot{\cal E}_1$ is the energy transfer rate from the mean flow, 
$\dot{\cal E}_2$ the viscous dissipation rate, 
$\dot{\cal E}_3$ the thermal diffusion rate and 
$\dot{\cal E}_4$ the shear-work rate, respectively; 
note that  the convective transfer of perturbation energy
by the mean flow, $\dot{\cal E}_0$, is zero.

\end{appendix}

\end{document}